\documentclass{emulateapj}
\usepackage{epsfig,graphics}

\newcommand{\si}{$\sigma$}
\newcommand{\be}{\begin{equation}}
\newcommand{\ee}{\end{equation}}

\newcommand{\msun}{{M$_{\odot}$}}
\newcommand{\mstar}{{M$_{\star}$}}
\newcommand{\ledd}{L$_{\rm Edd}$~}

\newcommand{\se}{s$^{-1}$ }
 
\newcommand{\gtsima}{$\; \buildrel > \over \sim \;$}
\newcommand{\ltsima}{$\; \buildrel < \over \sim \;$}
\newcommand{\prosima}{$\; \buildrel \propto \over \sim \;$}
\newcommand{\gsim}{\lower.5ex\hbox{\gtsima}}
\newcommand{\lsim}{\lower.5ex\hbox{\ltsima}}
\newcommand{\simgt}{\lower.5ex\hbox{\gtsima}}
\newcommand{\simlt}{\lower.5ex\hbox{\ltsima}}
\newcommand{\simpr}{\lower.5ex\hbox{\prosima}}

\newcommand{\esc}{erg~cm$^{-2}$~s$^{-1}$}

\newcommand{\etal}{{et al.~}}
\newcommand{\cxo}{\textit{Chandra~}}

\newcommand{\spi}{\textit{Spitzer~}}
\newcommand{\bhm}{M$_{\rm BH}$}
\newcommand{\lx}{L$_{\rm X}$}
\newcommand{\lb}{L$_{\rm B}$}
\shorttitle{AMUSE-Virgo I.: SMBHs in low-mass spheroids}
\shortauthors{Gallo et al.}

\begin{document}
\title{AMUSE-Virgo I. Super-massive black holes in low-mass spheroids}  
\normalsize
\author{Elena Gallo\altaffilmark{1,2}, Tommaso Treu\altaffilmark{1,3},
Jeremy Jacob\altaffilmark{1}, Jong-Hak
Woo\altaffilmark{1}, Philip J. Marshall\altaffilmark{1,4},  Robert
Antonucci\altaffilmark{1}} 
\altaffiltext{1}{Physics Department, University of
California Santa Barbara, CA 93106-9530} 
\altaffiltext{2}{Chandra Fellow}
\altaffiltext{3}{Sloan Fellow, Packard Fellow}
\altaffiltext{4}{Tabasgo Fellow}

\begin{abstract}
We present the first results from the AGN Multiwavelength Survey of Early-type
galaxies in the Virgo cluster (AMUSE-Virgo). This large program targets 100
early-type galaxies with the Advanced CCD Imaging Spectrometer on board the
{\it Chandra X-ray Observatory} and the Multi-band Imaging Photometer on board
the {\it Spitzer Space Telescope}, with the aim of providing an unbiased
census of low-level super-massive black hole activity in the local universe.
Here we report on the \cxo observations of the first 16 targets, and combine
them with results from archival data of another, typically more massive, 16
targets.  Point-like X-ray emission from a position coincident with the
optical nucleus is detected in 50 per cent of the galaxies (down to our
completeness limit of $\sim 4\times 10^{38}$ erg \se). Two of the X-ray nuclei
are hosted by galaxies (VCC1178=N4464 and VCC1297=N4486B) with absolute B
magnitudes fainter than $-$18, where nuclear star clusters are known to become
increasingly common. After carefully accounting for possible contamination
from low mass X-ray binaries, we argue that the detected nuclear X-ray sources
are most likely powered by low-level accretion on to a super-massive black
hole, with a $\simlt$11 per cent chance contamination in VCC1178, where a star
cluster is barely resolvable in archival {\it Hubble Space Telescope}
images. Based on black hole mass estimates from the global properties of the
host galaxies, all the detected nuclei are highly sub-Eddington, with
luminosities in the range $-8.4<$log(L$_{\rm 0.3-10~keV}/\rm L_{\rm
Edd}$)$<-5.9$.  The incidence of nuclear X-ray activity increases with the
stellar mass
\mstar\ of the host galaxy: only between 3--44 per cent of the galaxies with
\mstar$<10^{10}$
\msun\ harbor an X-ray active super-massive black hole. The fraction rises to between 49--87 per cent in
galaxies with stellar mass above $10^{10}$ \msun\ (at the 95 per cent
confidence level).

\end{abstract}

\keywords{black hole physics -- galaxies: nuclei -- galaxies: clusters:
individual (Virgo)}

\section{Introduction}

One of the main recent developments in the study of galaxy formation and
evolution has been the realization of the key role played by nuclear activity
due to accretion onto super-massive black holes (SMBHs).  Low-level
accretion-powered activity has been suggested to be relevant for a variety of
phenomena, including regulating star formation at galaxy scales via energy
feedback to solve the `downsizing' problem and providing extra energy to solve
the `cooling flow' problem (Cowie et al. 1996; Dalla Vecchia et al. 2004,
Springel \etal 2004, Treu et al. 2005a,b; Bundy et al. 2005, 2007; Juneau et
al. 2005; De Lucia et al. 2006, Abraham \etal 2007, Sijacki \etal 2007,
McNamara \& Nulsen 2007).  The most compelling pieces of evidence supporting a
strong connection between galaxy formation and nuclear activity are the tight
empirical scaling relations connecting the mass of the central SMBH with
global properties of the host galaxy, such as bulge luminosity and mass
(Kormendy \& Richstone 1995; McLure
\& Dunlop 2002; Marconi \& Hunt 2003; H\"aring \& Rix 2004), galaxy-light
concentration (Graham et al. 2001; Erwin 2004), and stellar velocity
dispersion (Gebhardt et al. 2000; Ferrarese \& Merritt 2000).  An issue of
fundamental importance in understanding the galaxy-black hole connection is
the `duty cycle' of nuclear activity, and its dependence upon, e.g. black hole
mass.  If SMBHs are indeed ubiquitous in galactic bulges, little is known about
the frequency and intensity of nuclear activity, the more so at the low-mass
end (see Greene \& Ho 2007a).  Even though a minimum level of accretion should
be present, fueled by mass loss during stellar evolution (e.g. Ciotti et
al. 1991; Ciotti \& Ostriker 2007), the inferred accretion-powered
luminosities are often lower than what expected from standard Bondi-Hoyle
accretion (as found e.g. for the Galactic Center SMBH Sgr A$^{\star}$;
Baganoff \etal 2003).

{From an empirical point of view, optical studies are mostly limited
to samples of known active nuclei (e.g. Woo \& Urry 2002; Heckman et
al. 2004; Kollmeier et al. 2006; Greene \& Ho 2007b) with limited
coverage of the black hole mass-Eddington luminosity (\bhm-L$_{\rm Edd}$) plane.}
Prior to the launch of the {\it Chandra X-ray Observatory}, searches
for low-level accretion powered X-ray emission from apparently
inactive galaxies were effectively limited to X-ray luminosities
$\simgt 10^{40}$ erg \se (e.g. Fabbiano \& Juda 1997; Allen, Di
Matteo \& Fabian 2000; Sulkanen \& Bregman 2001). The greatly improved
\cxo sensitivity, together with its fine spatial resolution, has made
it possible to investigate nuclear emission associated with SMBH
activity down to 3 orders of magnitude deeper, effectively bridging
the gap between active galactic nuclei (AGN) and inactive
galaxies. Perhaps surprisingly, only very low levels of nuclear X-ray
luminosity (\lx/\ledd$<10^{-6}$, 2-10 keV) have been observed in nearby {\it
massive} ellipticals (Di Matteo \etal 2000; Ho \etal 2001; Loewenstein
\etal 2001; Pellegrini 2005; Soria \etal 2006a,b; Santra \etal 2007), despite their
containing vast fuel reservoirs in the form of hot X-ray emitting interstellar
gas.  While these results rule out radiatively efficient solutions for the
accretion flow, the detected X-ray luminosities can vary by orders of
magnitude when plotted against the Bondi accretion rate (Pellegrini 2005), 
with a large fraction of systems being even fainter than
predicted by advection-dominated accretion flow models (Narayan \& Yi 1994).

These observations are closely related to the role of SMBH feedback in
inhibiting star formation at a galaxy scale level. Semi-analytical
models applied to state of the art cold-dark-matter simulations have
recently highlighted the importance of highly sub-Eddington SMBH
activity. In the formulation by Croton \etal (2006), a low level of
prolonged activity (the so-called `radio mode') is essential to
prevent the reservoir of gas surrounding the most massive galaxies
from cooling and producing young
stars, and thus reproduce their red colors.  In this scenario,
low-level SMBH feedback halts the gas supply to the disk from the
surrounding hot halo, truncating star formation and allowing the
existing stellar population to redden.

So far however, these studies have been sparse and necessarily focused on the
small number of galaxies at the high-mass end of the local population.  At the
same time, while the paucity of AGN in the local universe is a known
phenomenon, recent studies point towards an actual decline in the spatial
density of local active black holes with mass below $10^{6.5-7}$\msun\ (Greene
\& Ho 2007b), possibly due to low black hole occupation fraction and/or low bulge
fraction in dwarf galaxies.  In turn, this can place constraints on the very
mechanism by which SMBHs formed in the early universe, since different models
for the formation of black hole seeds predict different black hole occupation
fractions at redshift zero. This effect becomes more prominent down the mass
function. In particular, models where the black hole seeds are formed in the
nuclei of gravitationally unstable pre-galactic discs that form through the
collapse of haloes at redshift $\sim$10 (e.g., Madau \& Rees 2001, Begelman et
al. 2006, Lodato \& Natarajan 2006) predict the existence of a population of
faint low-mass galaxies with no black hole at their center (Volonteri \etal
2007a,b).

As a matter of fact, `light' SMBHs, presumably harbored by faint dwarf
galaxies, remain elusive: the strong limits placed by dynamical
studies on the masses of the nuclear objects in M33 (Gebhardt \etal
2001; Merritt \etal 2001) and NGC205 (Valluri \etal 2005) suggest that
neither galaxy hosts a SMBH of the mass expected from extrapolation of
the known scaling relations in massive bright galaxies.  Ferrarese
\etal (2006a) suggest that, while SMBHs are common in bright (absolute B
magnitude M$_{\rm B} <-$ 20) massive galaxies, they would be progressively
replaced by compact stellar nuclei moving down the mass function, and
may disappear entirely at the faint end. On the other hand, compelling
evidence exists for a $3.7\times 10^6$ \msun\ black hole at the center
of our own Milky Way (Ghez \etal 2005), providing us with the best
example of highly radiatively inefficient black hole accretion (the
measured X-ray luminosity between 2--10 keV can be as low as
$10^{33.3}$ erg \se; Baganoff \etal 2003).  Although no { direct 
dynamical} black hole mass determination exists below $10^6$ \msun, indirect
evidence points towards the existence of such objects in active
galaxies (Filippenko \& Ho 2003; Peterson \etal 2005; Barth \etal 2004;
Greene \& Ho 2004, 2007a), globular clusters (Gebhardt \etal 2002,
Gerssen \etal 2002) as well as (some) ultra-luminous X-ray sources
(Miller 2005).

{In this paper we present the first results from an extensive
multi-wavelength survey of 100 spheroids --elliptical, lenticular and
dwarf spheroidal galaxies-- in the Virgo cluster, conducted with the
\cxo {\it X-ray Observatory} and the {\it Spitzer Space Telescope}:
AMUSE-Virgo (AGN Multiwavelength Survey of Early-types in
Virgo\footnote{http://tartufo.physics.ucsb.edu/$^{\sim}$amuse}).  As described in
~\S~\ref{sec:amuse} the survey is designed to provide the first
unbiased census of low levels of nuclear activity in the local
universe as a function of host galaxy mass for early-type galaxies. \S~\ref{sec:data}
describes our analysis of the new and archival \cxo data obtained so
far (32/100 galaxies), as well as of archival {\it Hubble Space
Telescope} (HST) images used to connect the X-ray detections with their
optical counterparts and with the host galaxy properties. In
\S~\ref{sec:bhm} we use the known correlations with host galaxy
properties (stellar velocity dispersion $\sigma$ and spheroid
luminosity \lb) to estimate masses for the central black
holes. \S~\ref{sec:res} presents our main results, which are summarized in~\S~\ref{sec:sum}. }

\section{AMUSE-Virgo: program description}
\label{sec:amuse}

This \cxo Large Program (ID 08900784, Cycle 8, 454 ks; PI Treu) targets
the 100 early-type galaxies of the ACS\footnote{Advanced Camera for
Surveys, on board the {\em Hubble Space Telescope}.} Virgo Cluster
Survey (ACSVCS; C\^ot\'e \etal 2004), with the aim of providing an
unbiased census of SMBH luminosity in the local universe.
Mid-infrared observations with the Multi-band Imaging Photometer on
board {\it Spitzer} (MIPS; total exposure 9.5 hrs) complete the X-ray 
survey, allowing us to probe obscured accretion-powered emission
through 24 $\mu$m observations.

The \cxo survey has been designed to be sensitive (at 3-$\sigma$) to a
3 \msun\ object accreting at the Eddington limit.  As described in
detail in \S~\ref{sec:res}, this is the optimal depth for an extensive
survey: the threshold is deep enough to be interesting, yet bright
enough to ensure negligible contamination by stellar mass black holes
(or background sources) within the Chandra Point Spread Function (PSF). The
desired sensitivity is accomplished by means of snapshot (5.4 ksec)
observations of 84 targets. The new data are combined with deeper
archival \cxo observations of the remaining (on average more massive)
16 targets. 

Based on the comparison with the spectral energy distribution (SED) of LINERs
(Low Ionization Nuclear Emitting Regions; see e.g. Maoz \etal 1998) and
un-obscured AGN (both radio loud and radio quiet), the mid-IR band flux is
expected to exceed the \cxo flux by at least a factor 3.  Hence, the \spi
survey -- which will acquire new data for 57 objects, to be combined with
archival data for the remaining 43 -- has been designed to probe down to
$\simeq 3\times 10^{-14}$ erg sec cm$^{-2}$ (three times higher than the \cxo
threshold).

Based on empirical scaling relations between the black hole mass and
the host properties, the ACSVCS sample covers over 5 orders of
magnitude in black hole mass as estimated from the mass-velocity
dispersion
relation (see~\S~3 for a critical assessment), large enough that it can be
divided in SMBH mass bins to test whether the nuclear activity duty
cycle is mass dependent; given our sensitivity, we will probe X-ray
Eddington ratios in the range $10^{-9}-10^{-5}$.

\section{Data analysis}
\label{sec:data}

In this section we report on the analysis of \cxo data for the 16
targets observed in Cycle 8 at the time of submission of this paper (\S~\ref{sec:c8}), and on the analysis
of the nuclear X-ray emission of the 16 galaxies of the survey that
have archival \cxo data (\S~\ref{ssec:car}). \S~\ref{ssec:HST}
describes the analysis of archival HST data used to
compare the location of X-ray detections to the optical center of the
host galaxies (and nuclei when present), and to estimate their stellar
mass.  The target list and observation log are given in Table~\ref{tab:log}. 
\begin{figure}
\center{
\includegraphics[angle=0,scale=.47]{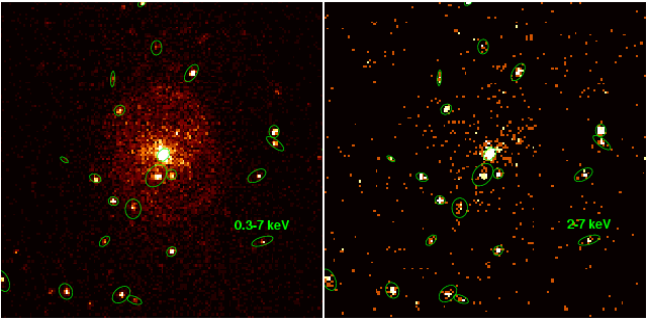}
\caption{ 
ACIS-S image of VCC1632 (=M89; 24.8 ksec of exposure). {\em Left:}
0.3--7 keV image. {\em Right:} 2--7 keV image, with negligible diffuse
emission from the hot gas. 
\label{fig:m89}}}
\end{figure}
\subsection{\cxo Cycle 8 data}
\label{sec:c8}

We observed each galaxy with the Advanced CCD Imaging Spectrometer
(ACIS) detector on board \cxo for 5.4 ksec of nominal exposure time in
Faint-mode. The target was placed at the aim point of the back-side
illuminated S3 chip.  We analyzed the data using the \cxo Interactive
Analysis Observation (CIAO) software version 3.4.1.1 and the
calibration database version 3.3.0.1. Standard level 2 event lists,
processed for cosmic ray rejection and good time filtering, were
employed.  As \cxo is known to encounter periods of high background
which especially affect the S1 and S3 chips, we first checked for
background flares and removed time intervals with background rate
$\simgt 3\sigma$ above the mean level.

Further analysis was restricted to energies between 0.3--7.0 keV in order to avoid
calibration uncertainties at low energies and to limit background
contaminations at high energies.
We applied a wavelet detection algorithm over each activated chip,
using CIAO {\tt wavdetect} with sensitivity threshold corresponding 
to a $10^{-6}$ chance of detecting one spurious source per PSF element if the local background is uniformly
distributed.
We used the default wavelet parameters, with scales increasing by a
factor of $\sqrt{2}$: between 1--4 pixels on a full resolution
circular region of 512 pixel radius centered on the nominal position
of the target (restricting the circle to the sole S3 chip), and
between 1--8 pixels for a 1\arcsec~resolution image of each activated
chip.  The complete \cxo source catalog will be presented elsewhere; here
we shall focus on the X-ray properties of the nuclei. 
Given the relatively short exposures ($\sim5.4$ ksec), we detect between 3
(VCC33) and 15 (VCC751) X-ray sources within a 512 pixel ($\sim$252\arcsec)
aperture. A small fraction of them, often none, are enclosed
within the HST ACS field of view ($200\times200$ arcsec$^2$); this prevents us
from directly registering the \cxo data to HST.
Thus, we first improved the \cxo astrometry by cross-matching
the detected (non-nuclear) X-ray sources with the Sloan Digital Sky Survey (SDSS)
catalog (Data Release 5, DR5), and applied the resulting bore-sight
corrections following the method described by Zhao \etal (2005). 
\begin{center}
\newcommand\tabspace{\noalign{\vspace*{0.7mm}}}
\def\errtwo#1#2#3{$#1^{+#2}_{-#3}$}
\begin{deluxetable*}{lllllllllll} 
\setlength{\tabcolsep}{0.005in} 
\tabletypesize{\scriptsize} 
\tablewidth{0pt} 
\tablecaption{AMUSE-Virgo I.: Observation log and X-ray nuclei \label{tab:log}}
\tablehead{      \colhead{ID}
	   &     \colhead{VCC}
	   &     \colhead{Other}
	   &     \colhead{OID}
	   &     \colhead{Date}
	   &     \colhead{Exp.}   	   
	   &     \colhead{X-ray}
           &     \colhead{ } 
           &     \colhead{ }
	    &     \colhead{Opt Nucl.}
	    &     \colhead{}	   
\\	 
		\colhead{}          
           &       \colhead{}
  	   &       \colhead{}
           &     \colhead{}
	   &     \colhead{ }
  	   &     \colhead{ }	   
	   &     \colhead{Counts}
           &     \colhead{$\alpha$(J2000)}
	   &     \colhead{$\delta$(J2000)}
           &     \colhead{$\alpha$(J2000)}
	   &     \colhead{$\delta$(J2000)} 
           \\	
	         \colhead{(1)} 
	   &	 \colhead{(2)} 
           &     \colhead{(3)} 
           &     \colhead{(4)}  
	   &     \colhead{(5)}  
	   &     \colhead{(6)}  
	   &     \colhead{(7)} 
	   &     \colhead{(8)} 
	   &     \colhead{(9)} 
	   &     \colhead{(10)} 
	   &     \colhead{(11)} 
} 
\startdata 
	    1
	    & 1226
	    & N4472 
            & 321$^{\rm a}$
	    & 00-06-12
	    & 33.64
	    & ..
	    & ..
	    & ..
	    & 12:29:46.764
	    & +08:00:01.75
	    \\
	    2
	    & 1316
	    & N4486 
            & 352,  
	    & 00-07-29, 
	    & 119.22
            & 82011$^{\rm b}$ (961)
	    & 12:30:49.409 (0.01)
	    & +12:23:28.20 (0.01)
            & 12:30:49.425
            & +12:23:28.08      
	\\
	    & 
	    & 
            & 2707,
	    & 02-07-06,
	    &
            &  
	    & 
	    & 
            &
            &              
	\\
	    & 
	    & 
            & 3717
	    & 02-07-05
	    & 
            &  
	    & 
	    & 
            &
            &      
	\\
	    3
	    & 1978
	    & N4649 
            & 785$^{\rm a}$
	    & 00-04-20
	    & 34.31
	    & 151.3$^{\rm b}$ (61.2)
	    & 12:43:39.977 (0.02)
	    & +11:33:10.05 (0.02)
	    & 12:43:39.989
	    & +11:33:09.55	    	    
\\
	    4
	    & 881
	    & N4406 
            & 318
	    & 00-04-07
	    & 14.63
	    & ..
	    & ..
	    & ..
	    & 12:26:11.759 
	    & +12:56:46.44	    	    
\\
		5
	    & 798
	    & N4382 
            & 2016
	    & 01-05-29 
	    & 39.75
	    & ..
	    & ..
	    & ..
	    & 12:25:24.071
	    & +18:11:27.92
\\
		6
	    & 763
	    & N4374 
            & 6131$^{\rm a}$
	    & 05-11-07
	    & 38.75
	    & 705$^{\rm b}$ (63) 
	    & 12:25:03.746 (0.04)
	    & +12:53:13.30 (0.04)
	    & 12:25:03.743
	    & +12:53:13.01
\\
	    7
	    & 731
	    & N4365
            & 5921$^{\rm a}$
	    & 05-04-28
	    & 39.51
	    & 84.8$^{\rm b}$ (27) 
	    & 12:24:28.289 (0.06)
	    & +07:19:04.03 (0.06)
	    & 12:24:28.271
	    & +07:19:03.99
\\
		8 
	    & 1535
	    & N4526
            & 3925$^{\rm a}$
	    & 03-11-14 
	    & 38.39
	    & ..
	    & ..
	    & ..
	    & 12:34:03.029
	    & +07:41:57.62
\\
		9
	    & 1903 
	    & N4621 
            & 2068
	    & 01-08-01
	    & 24.84
	    & 167.4$^{\rm b}$ (30.5) 
	    & 12:42:02.242 (0.06)
	    & +11:38:49.17 (0.06)
	    & 12:42:02.262
	    & +11:38:48.87
\\
		10
	    & 1632
	    & N4552 
            & 2072$^{\rm a}$
	    & 01-04-22
	    & 54.13
	    & 933$^{\rm b}$ (82) 
	    & 12:35:39.811 (0.02)
	    & +12:33:22.81 (0.02)
	    & 12:35:39.811
	    & +12:33:22.74
\\	    
		11
	    & 1231
	    & N4473
            & 4688$^{\rm a}$
	    & 05-02-26
	    & 29.58
	    & 58.7$^{\rm b}$ (13.6) 
	    & 12:29:48.864 (0.12)
	    & +13:25:45.19 (0.17)
	    & 12:29:48.870
	    & +13:25:45.95
\\
		12
	    & 2095
	    & N4762
            & 3998$^{\rm a}$
	    & 03-05-17 
	    & 5.31 
	    & 11.8 (3.5)
	    & 12:52:55.979 (0.10)
	    & +11:13:51.63 (0.09)
	    & 12:52:56.045 
	    & +11:13:51.87
\\
		13
 	    & 1154
	    & N4459
	    & 2927
            & 02-06-02 
	    & 9.83
	    & 48.3 (7)
	    & 12:29:00.027 (0.13)
	    & +13:58:42.56 (0.12)
	    & 12:29:00.030
	    & +13:58:42.90
\\
		15
	    & 2092 
	    & N4754
	    & 8038
            & 07-02-19
	    & 4.97
	    & 8.7 (2.9)
	    & 12:52:17.496 (0.14)
	    & +11:18:49.99 (0.19)
	    & 12:52:17.50
	    & +11:18:49.97
	    \\
	     18     
	    & 1692 
	    & N4570
	    & 8041
            & 07-04-28
	    & 5.09
	    & 5.7 (2.4)
	    & 12:36:53.378 (0.14)
	    & +07:14:47.98 (0.22) 
	    & 12:36:53.398
	    & +07:14:47.57
\\
	    21
	    & 685 
	    & N4350
            & 4015
            & 04-02-28
	    & 4.7
	    & 27.8 (5.3)
	    & 12:23:57.869 (0.15)
	    & +16:41:36.42 (0.17)
	    & 12:23:57.848
	    & +16:41:36.21
	\\
	     22
	    & 1664
	    & N4564
            & 4008$^{\rm a}$
            & 03-11-21 
	    & 17.86
	    & 74.6$^{\rm b}$ (9)
	    & 12:36:26.990 (0.11)
	    & +11:26:21.70 (0.10)
	    & 12:36:26.995
	    & +11:26:21.04
	\\
	     27
	    & 1720 
	    & N4578
            & 8048
            & 07-04-28
	    & 5.09 
	    & ..
	    & ..
	    & .. 
	    & 12:37:30.561
	    & +09:33:18.19
\\
	    30
	    & 1883 
	    & N4612
            & 8051
            & 07-04-16
	    & 5.09 
	    & 4.8 (2.2)
	    & 12:41:32.743 (0.18)
	    & +07:18:52.93 (0.26)
	    & 12:41:32.751
	    & +07:18:53.48
\\
	    42
	    & 1913 
	    & N4623
            & 8062
            & 07-05-11
	    & 5.28
	    & ..
	    & ..
	    & .. 
	    & 12:42:10.688
	    & +11:38:48.87
\\
		46 
	    & 1178 
	    & N4464
            & 8127
            & 07-04-29
	    & 5.09 
	    & 10.8 (3.3)
	    & 12:29:21.297 (0.21)
	    & +08:09:23.80 (0.17)
	    & 12:29:21.289 
	    & +08:09:23.81
\\
	    51
	    & 2048 
	    & I3773
            & 8070
            & 07-04-28
	    & 5.28
	    & ..
	    & ..
	    & .. 
	    & 12:47:15.293
	    & +10:12:12.82
\\
	     53
	    & 9 
	    & I3019
            & 8072
            & 07-04-13
	    & 5.28
	    & ..
	    & ..
	    & .. 
	    & 12:09:22.340
	    & +13:59:33.10
\\
		56
	    & 1049 
	    & U7580
            & 8075
            & 07-04-28
	    & 5.51
	    & ..
	    & ..
	    & .. 
	    & 12:27:54.836
	    & +08:05:25.40
\\
	     61
	    & 1297
	    & N4486B 
            & 4007$^{\rm a}$
            & 03-11-21
	    & 36.18 
	    & 42.2$^{\rm b}$ (8.2)
	    & 12:30:31.966 (0.08)
	    & +12:29:24.64 (0.07)
	    & 12:30:31.969
	    & +12:29:24.57
\\
	    67
	    & 1833 
	    & $-$
            & 8084
            & 07-03-17
	    & 5.25
	    & ..
	    & ..
	    & .. 
	    & 12:40:19.670
	    & +15:56:06.76 
\\
	    70
	    &  33
	    & I3032
            & 8086
            & 07-02-01
	    & 5.16
	    & ..
	    & ..
	    & .. 
	    & 12:11:07.755 
	    & +14:16:29.20
\\
	    76
	    & 1895 
	    & U7854
            & 8092
            & 07-04-28
	    & 5.09
	    & ..
	    & ..
	    & .. 
	    & 12:41:51.964
	    & +09:24:10.33
\\
	    80
	    & 1857 
	    & I3647
            & 8130
            & 07-04-29
	    & 5.09
	    & ..
	    & ..
	    & .. 
	    & 12:40:53.095
	    & +10:28:33.46
\\
	    86
	    & 2050 
	    & I3779
            & 8101
            & 07-05-07
	    & 5.09
	    & ..
	    & ..
	    & .. 
	    & 12:47:20.638
	    & +12:09:59.12
\\
	    88
	    & 751 
	    & I3292
            & 8103
            & 07-04-13 
	    & 5.09
	    & ..
	    & ..
	    & .. 
	    & 12:24:48.363
	    & +18:11:42.44 
\\
	    93
	    & 1199 
	    & $-$
            & 8107
            & 07-04-28
	    & 5.09
	    & ..
	    & ..
	    & .. 
	    & 12:29:34.998
	    & +08:03:28.81
\\
\\
\tabspace 
\enddata 
\tablecomments{Col.: (1) ACSVCS target number (2) VCC source name; (3)
Other name (4) \cxo\ Observation Identity; (5) Observation starting date; (6)
Net exposure (after flares' removal) (7) Nuclear X-ray source: net counts
extracted --or extrapolated to-- between 0.3--7 keV; (8) X-ray Nucleus
R.A. (J2000), with the positional uncertainty on the centroid position given
in parenthesis, in arcsec (9) X-ray Nucleus Dec (J2000) with the positional uncertainty
on the centroid position given in parenthesis, in arcsec; (10) Optical nucleus
R.A. (J2000); (11) Optical nucleus Dec (J2000). \\ $\rm a)$: Very Faint mode;
$\rm b)$ In order to avoid contamination from the diffuse gas emission, the
nuclear counts were extracted between $E_{t}$-7 keV (see \S~\ref{ssec:car}),
where typically $E_{t}\simgt 2$ keV, and then extrapolated to between 0.3--7
keV within {\tt webPimms} adopting an absorbed power law model with Galactic
absorption and photon index $\Gamma=2$. \label{tab:log}}
\end{deluxetable*} 
\end{center}
%
Individual source locations are subject to statistical uncertainties affecting
the centroiding algorithm and to the dispersion of photons due to the
PSF. For ACIS-S, Garmire et al. (2000) estimate 90 per cent
confidences of $\pm$0\arcsec\.5 for sources with $\sim$10 counts,
$\pm$0\arcsec\.2 for 20-50 count sources, and negligible for $>$100 count
sources. In addition, the statistical uncertainties depend on the off-axis
angle from the aim point: we calculated the 95 per cent confidence error radii, $r_{\rm X}$,
as a function of net counts and off-axis angle according to the empirical
formula based on the results of Hong et al. (2005). 
The statistical uncertainties affecting the centroid errors in the
positions of the X-ray sources, 
combined with the $\simlt$0\arcsec.1 positional error of SDSS, results
in a final astrometric frame that is accurate to between 0\arcsec.2
(fields with $\simgt$20 counts sources) and 0\arcsec.5
(fields with faint sources).

After registering the \cxo images to SDSS, we ran again {\tt 
wavdetect} to refine the positions, and searched for point-like X-ray
emission centered at the galaxy optical center, derived from archival
HST ACS images registered to the SDSS world coordinate
system as described in the next subsection.
We searched for X-ray counterparts to the ACS nuclei within an error circle
which is the quadratic sum of the positional uncertainty for the X-ray source,
the uncertainty in the optical astrometry, and the
uncertainty in the X-ray bore-sight correction, multiplied by the chosen
confidence level scale factor (3$\sigma$):
\begin{equation}
R_{\rm err}=\sqrt{r_{\rm X}^2+r_{\rm opt}^2+r_{\rm bore}^2}
\end{equation}
The coordinates of the detected X-ray nuclei, with their statistical
uncertainty, are listed in Table~\ref{tab:log}.  More details about the optical
astrometry are provided in \S~\ref{ssec:HST}.

For the X-ray aperture photometry, we
adopted a circular region with a 2\arcsec\ radius centered on X-ray centroid
position.
For all the observations
considered here, the aim-points were specified in order to have optimal nuclear
positions with respect to CHIP geometry and telescope focus; thus, 
2\arcsec\ correspond to 95 per cent of the encircled
energy radius at 1.5 keV for ACIS. 
We inspectioned the morphology of the detected nuclei by constructing the \cxo PSF
at 1.5 keV and normalized it to the actual number of detected counts; all the
detected X-ray nuclei in the Cycle 8 observations are consistent with being point-like. 
%
\begin{center}
\newcommand\tabspace{\noalign{\vspace*{0.7mm}}}
\def\errtwo#1#2#3{$#1^{+#2}_{-#3}$}
\begin{deluxetable*}{rrrclcccclc} 
\setlength{\tabcolsep}{0.05in} 
\tabletypesize{\scriptsize} 
\tablewidth{0pt} 
\tablecaption{AMUSE-Virgo I.: nuclear properties\label{tab:cxo}}
\tablehead{      \colhead{ID}
	   &     \colhead{VCC}
	   &     \colhead{Other}
	   &     \colhead{$d$}
	   &     \colhead{B}
	   &     \colhead{\si}
	   &     \colhead{log M$_{\rm BH_{B}}$}
	   &     \colhead{log M$_{\rm BH_{\sigma}}$}
  	   &     \colhead{log $\rm L_{\rm X,nucl}$}
           &    \colhead{log M$_{\star}$}
\\	 
		\colhead{}          
           &       \colhead{}
 	  &       \colhead{}
           &     \colhead{(Mpc)}
	   &     \colhead{(mag)}
  	   &     \colhead{(km/\se)}
           &     \colhead{(\msun)}
	   &     \colhead{(\msun)}
	   &     \colhead{(erg s$^{-1}$)}
           \\	
	         \colhead{(1)} 
	   &	 \colhead{(2)} 
           &     \colhead{(3)} 
           &     \colhead{(4)}  
	   &     \colhead{(5)}  
	   &     \colhead{(6)}  
	   &     \colhead{(7)} 
	   &     \colhead{(8)} 
	   &     \colhead{(9)} 
	   &     \colhead{(10)} 
} 
\startdata 
	     1
	    & 1226$^a$ 
	    & M49, N4472 
            & 17.14 
	    & 8.63
	    & 308$\pm$9   
	    & 9.4 
	    & 9.1 
	    & $<$38.49 
            (14,15)
  	    & 12.0
	     \\
		2
	    & 1316$^a$ 
	    & M87, N4486
            & 17.22 
	    & 9.05
	    & 355$\pm$8   & 9.2 & 9.4
	    & 41.20 
	    (16,17)
	 & 11.8
	\\
		3
	    & 1978$^a$ 
	    & M60, N4649
            & 17.30 
	    & 9.33
	    & 347$\pm$9   & 9.1 & 9.4 
	    & 39.05 
	    (18) 
	     & 11.7
\\
		4
	    & 881$^a$ 
	    & M86, N4406
            & 16.83 
	    & 8.77
	    & 245$\pm$11   & 9.3 & 8.6
	    & $<$38.64
	    & 11.9
\\
		5
	    & 798$^a$ 
	    & M85, N4382
            & 17.86 
	    & 9.30
	    & 205$\pm$8   & 9.2 & 8.3
	    & $<$38.43 
	    (19)
	    & 11.6
\\
		6
	    & 763$^a$ 
	    & M84, N4374
            & 18.45 
	    & 9.35
	    & 297$\pm$7   & 9.2 & 9.0
	    & 39.73
	    (20) 
	    & 11.7
\\
	    7
	    & 731$^a$ 
	    & N4365
            & 23.33
	    & 9.98
            & 261$\pm$7   & 9.1 & 8.8 
	    & 39.0 
	    (19)
	& 11.7
\\
		8 
	    & 1535$^a$ 
	    & N4526
            & 16.50 
	    & 10.52$^b$
	    & 316$\pm$7   & 8.6 & 9.2
	    & $<$38.21
	    & 11.0
\\
		9
	    & 1903$^a$ 
	    & M59, N4621
            & 14.93 
	    & 10.02
	    & 233$\pm$7   & 8.7 & 8.5 
	    & 39.11
	& 11.3
\\
		10
	    & 1632$^a$ 
	    & M89, N4552
            & 15.85 
	    & 10.13
	    & 257$\pm$18   & 8.7 & 8.7
	    & 39.58 
	    (21) 
	& 11.3
\\	    
		11
	    & 1231$^a$ 
	    & N4473
            & 15.27 
	    & 11.19
	    & 189$\pm$10   & 8.2 & 8.1 
	    & 38.60$^c$
	    & 10.8
\\
		12
	    & 2095$^a$ 
	    & N4762
            & 16.50 
	    & 11.97
	    & 147$\pm$10 (12) & 8.0 & 7.6
	    & 38.71 
	    & 10.6
\\
		13
 	    & 1154$^a$ 
	    & N4459
            & 16.07 
            & 11.07
	    & 170$\pm$12   & 8.3 & 7.9
	    & 39.03
	    (22) 
	    & 10.9
\\
		15
	    & 2092 
	    & N4754
            & 16.14
	    & 11.36
 	    & 200$\pm$10 (12) & 8.2 & 8.2
	    & 38.59 
	    & 10.9
	    \\
	     18     
	    & 1692 
	    & N4570
            & 17.06
            & 11.91
	    & 180$\pm$18   & 8.0 & 8.0
	    & 38.45
	    & 10.6
\\
	    21
	    & 685$^{a}$ 
	    & N4350
            & 16.50
	    & 11.83
	    & 198$\pm$9   & 8.0 & 8.2			
	    & 39.14
	    (23)
		& 10.6
	\\
	     22
	    & 1664$^{a}$ 
	    & N4564
            & 15.85
	    & 11.85
	    & 157$\pm$9   & 8.0 & 7.7
	    & 39.95
	    (24)
	    & 10.6
	\\
	     27
	    & 1720 
	    & N4578
            & 16.29
            & 12.01
	    & 153$\pm$15   & 8.0 & 7.6
	    & $<$38.54
	& 10.4
\\
	    30
	    & 1883 
	    & N4612
            & 16.59
	    & 12.01
	    & 104$\pm$11 (12) & 8.0 & 6.8
	    & 38.35
	    & 10.4
\\
	    42
	    & 1913 
	    & N4623
            & 17.38
	    & 13.16
	    & 89$\pm$10 (12) & 7.5 & 6.5
	    & $<$38.46
	    & 10.1
\\
		46 
	    & 1178 
	    & N4464
            & 15.85
	    & 13.32
	    & 121$\pm$25 (13) & 7.4 & 7.1
	    & 38.67
	    & 9.9
\\
	    51
	    & 2048 
	    & I3773
            & 16.50
	    & 14.04
	    & 79$\pm$5   & 7.1 & 6.3
	    & $<$38.12
	    & 9.5
\\
	     53
	    & 9 
	    & I3019
            & 17.14
	    & 14.00
	    & .. & 7.2 & ..
	    & $<$38.15
	     & 9.7
\\
		56
	    & 1049 
	    & U7580
            & 15.99
	    & 14.93
	    & .. & 6.7 & .. 
	    & $<$38.08
	    & 9.0
\\
	     61
	    & 1297$^a$ 
	    & N4486B 
            & 16.29
	    & 14.14
	    & 166$\pm$8   & 7.1 & 7.8 
	    & 38.42
	    (24)
	    & 9.7
\\
	    67
	    & 1833 
	    & $-$
            & 16.22
	    & 14.66
	    & .. & 6.8 & ..
	    & $<$38.11
	    & 9.3
\\
	    70
	    &  33 
	    & I3032
            & 15.07
	    & 15.22
	    & .. 
	    & 6.5 
	    & .. 
	    & $<$38.25
	    & 8.9
\\
	    76
	    & 1895 
	    & U7854
            & 15.85
	    & 15.15
	    & .. 
	    & 6.6 
	    & .. 
	    & $<$38.10
	    & 9.0
\\
	    80
	    & 1857 
	    & I3647
            & 16.50
	    & 15.06
	    & .. 
	    & 6.7 
	    & ..
	    & $<$38.14
	    & 9.4
\\
	    86
	    & 2050 
	    & I3779 
            & 15.78
	    & 15.37
	    & .. 
	    & 6.5 
	    & ..
	    & $<$38.10
	    & 9.0
\\
	    88
	    & 751 
	    & I3292
            & 15.78
	    & 14.86
	    & .. 
	    & 6.7 
	    & .. 
	    & $<$38.29
	    & 9.4
\\
	    93
	    & 1199 
	    & $-$
            & 16.50
	    & 16.00
	    & .. 
	    & 6.3 
	    & ..
	    & $<$38.14
	    & 9.0
\\
\\
\tabspace 
\enddata 
\tablecomments{Col.: (1) ACSVCS target number (2) VCC source name; (3)
Alternate name, from NCG or catalogs; (4)
Distance (from surface brightness fluctuations method; Mei \etal
2007). The average distance to the Virgo cluster -- of 16.5 Mpc -- is
employed in case of no available distance modulus; (5)
Extinction-corrected B magnitude, estimated as described in
\S~\ref{ssec:HST}; E(B-V) values are from Ferrarese \etal (2006b);
(6) Stellar velocity dispersion, from ENEARc (Bernardi \etal 2002), unless
otherwise indicated; (7) B-based black hole mass, adopting the scaling by FF05
(8) \si-based black hole mass, adopting the scaling by FF05;
(9) Nuclear luminosity between
0.3-10 keV, corrected for absorption; literature references are given in
brackets; (10) Stellar mass of the host galaxy, in \msun, derived from $g_0$ and
$z_0$ band AB model magnitudes following Bell \etal (2003), as
described in\S~\ref{ssec:HST}.\\ References 11)
Gavazzi
\etal 1999 (Ga99); 13) Davies \etal 1987 (D87); 14) Biller \etal 2004;
15) Soldatenkov et al. 2003; 16) Di Matteo et al. 2003; 17) Wilson
\& Yang 2002; 18) Randall \etal 2004 report on a point-like X-ray source at 1\arcsec.6
from the nucleus. After cross-matching the \cxo astrometry to SDSS, the
position of the nuclear X-ray source is found to be consistent with the galaxy
nucleus; 
19) Sivakoff \etal 2003; 20) Finoguenov \& Jones 2001; 21) Xu et
al. 2005; 22) Satyapal \etal 2005 ; 23) Dudik \etal 2005; 24) Soria
\etal 2006a,b; the nuclear X-ray source in VCC1297 has a soft
spectrum. However, a thermal model provides a statistically worse fit
that a power law model (employing Cash statistics), implying that most
of the X-ray emission is likely accretion-powered.\\ $a)$ Archival
data; $b)$ From B$_{\rm T}$ magnitude in Ferrarese \etal (2006b); $c)$ The
X-ray source appears slightly elongated; $\rm L_{\rm X,nucl}$ is estimated
within the \cxo PSF at 1.5 keV.
\label{tab:cxo}  }
\end{deluxetable*} 
\end{center}
We adopted an annulus with inner radius 20\arcsec\ and outer radius
30\arcsec\ for background subtraction (off-nuclear X-ray sources, if
present, were masked out). 
We estimated the corresponding fluxes using
{\tt webPimms}\footnote{http://heasarc.gsfc.nasa.gov/Tools/w3pimms.html},
and assuming an absorbed power-law model with photon index $\Gamma=2$
and hydrogen equivalent column $N_{\rm H}$=$2.5\times 10^{20}$
cm$^{-2}$, i.e. the nominal Galactic value determined from the HI studies of 
Dickey \& Lockman (1990). 
Since none of galaxies under exam shows evidence for prominent dusty lanes in
the Hubble images (Ferrarese \etal 2006b), it is reasonable to assume that the
Galactic value provides a correct estimate for the actual absorbing column. 
Under this assumptions, $10^{-3}$ count \se\ in the 0.3-7 keV energy band
correspond to an intrinsic flux of 7.19$\times 10^{-15}$ \esc\ between 0.3-10
keV (ACIS-S). 

\begin{figure}
\center{
\includegraphics[angle=0,scale=.47]{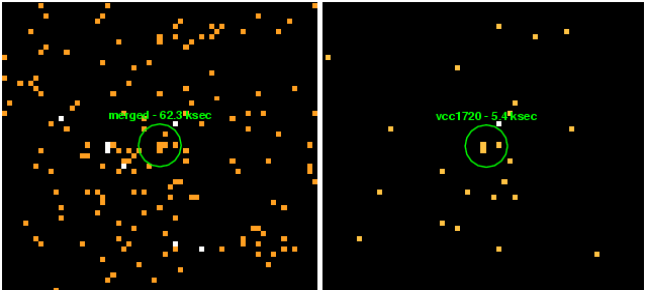}
\caption{ 
The stacked ACIS-S image (left panel) of the 12 out of 16 galaxies which
have snapshot (5.4 ksec) observation and no nuclear detection is
compared to the image of a single snapshot (VCC1720, right panel).  5
photons are detected within a 2\arcsec\ radius circular aperture
center on the stacked nuclear position (green circle), while 3.1
photons are expected from the background over the same area; this
implies no significant detection. 
\label{fig:stack}}}
\end{figure}

In case of no significant detection we applied Poisson statistics to derive
upper limits on the nuclear luminosity at the 95 per cent confidence level
(Gehrels 1986), listed in Table~\ref{tab:cxo}. To obtain a more stringent
limit on the average flux, we stacked the images of the non-detections
centered on the optical centers, resulting in 62.3 ks of effective total
exposure.  We extracted the counts from a 2\arcsec\ radius circular aperture,
and background from an annulus with inner and outer radii $R_{\rm
in}=$2\arcsec\ and $R_{\rm out}$=9\arcsec, centered on the stacked
nucleus position (see Figure~\ref{fig:stack}).  We found 5 counts within the
2\arcsec\ radius aperture, while 3.1 are expected from the background.  The
Poisson probability of obtaining 5 counts or more when 3.1 are expected is
0.2, indicating no significant detection. This corresponds to an {exposure
weighted, {\it average} count rate} $<1.6\times 10^{-4}$ count
\se\ for the undetected sources (95 per cent confidence level), or
$\langle$\lx$\rangle <3.8\times 10^{37}$ erg
\se\ (0.3--10 keV) at the average distance of 16.5 Mpc (Mei \etal 2007).

\subsection{\cxo archival data}
\label{ssec:car}

We followed the same procedure as outlined above for the 16 galaxies
which have \cxo archival data (marked by an {\it a} superscript in Table~\ref{tab:cxo}). For
these targets, event 1 lists were first filtered (and cleaned, in the
case of Vary Faint telemetry) following the standard CIAO threads. As
the archival sample is mainly made of massive, X-ray bright galaxies,
we had to model/account for the diffuse gas contribution in order to
constrain any possible accretion-powered, nuclear X-ray emission. This
was achieved by first determining for each galaxy the energy $E_t$ 
above which hot gas contribution is negligible.
The threshold energy $E_t$ was derived as follows. As a first
step, we extracted the spectrum of the total diffuse emission over a circular
aperture of 150\arcsec\ centered on the galaxy nucleus and 
excluding all the resolved point sources detected by {\tt 
wavdetect}. The background for this spectrum was extracted on the S3
chip as far away as possible from the galaxy, using a annulus of inner and
outer radius 250\arcsec and 300\arcsec, respectively (masking out the resolved
X-ray sources). 
We analyzed the extracted spectra with XSPEC version 11.2.0
(Arnaud 1996), using a combination of optically thin thermal emission 
for the diffuse gas plus a non-thermal component (power-law model) to represent the emission from the
unresolved point sources, under the assumption that the hard spectral
component seen in the diffuse emission is mainly due to the
contribution of unresolved low-mass X-ray binaries (LMXBs). We fixed
the power-law photon index $\Gamma$ of the hard component due to
unresolved LMXBs to the value measured for the cumulative spectrum of
all the resolved X-ray sources ($\Gamma=$ 1.6--1.9).
As a model for the diffuse thermal emission, we employed the 
Astrophysical Plasma Emission Code (APEC) thermal-emission
model (Smith et al. 2001) in its most recent version ({\tt vapec}), which
includes a wealth of accurate atomic data. The abundances of neon, magnesium,
silicon, and iron were left free to vary. The two 
spectral components are subjected to a common absorption 
($N_H=2.5\times 10^{20}$ cm$^{-2}$). As a consistency-check, we also re-run
the fits by letting both the $N_H$ column and the power law photon index vary,
and recovered the same parameters, within errors.

The fits yielded the temperature of the hot thermally-emitting gas, $kT_g$,
for each galaxy, and allowed us to estimate the energy $E_t$ above which 
the optically thin thermal emission contributes to less than 5 per cent to the
measured flux. 
As an example, the
0.3-7 keV spectrum of the diffuse emission of VCC1978(=N4649) is best-fit by a
two component model with total flux of $2.99 \times 10^{-12}$ \esc. ~The absorbed thermal
component accounts for 69 per cent of this flux, but contributes to less than
5 per cent above $E_t=1.92$ keV. In this case, the fitted gas temperature is $kT_g=0.78_{-0.02}^{+0.01}$ KeV.
As in the analysis of the new data described in~\S~\ref{sec:c8}, we converted
the measured count rates ($E_t$-7 keV) extracted from a 2\arcsec~circular
region centered at the optical nucleus into 0.3-10 keV unabsorbed luminosity
within {\tt webPimms}, by assuming an absorbed power law model with photon
index $\Gamma=2$.  We extracted and fitted the actual source
spectrum when dealing with more than 50 hard ($>E_t$) photons (i.e. for
VCC1613=M87, VCC1632=N4552, obtaining $\Gamma=2.21\pm0.04$,
$\Gamma=1.7^{+0.8}_{-0.5}$ respectively).
\subsection{{Hubble Space Telescope} imaging} 

\label{ssec:HST}

Images from the ACS Virgo Cluster Survey (C\^ot\`e
\etal 2004) were downloaded from the HST archive. The 
observations of each galaxy consist of two 375s exposures in the F475W filter
(nearly equivalent to SDSS $g'$-band, effective wavelength $\lambda_{\rm
eff}$=4825~\AA; Fukugita \etal 1996), two 560s exposures in the F850LP filter (nearly
equivalent to SDSS $z'$-band, $\lambda_{\rm eff}$=9097~\AA; Fukugita \etal 1996), and
a single 90s exposure in the F850LP filter, all in the wide field channel.  In
order to facilitate the best possible matching between X-ray and optical
sources, the astrometry of HST images has been referenced to the that of
SDSS-DR5 according to the following procedure. First, individual exposures in
each band are combined using the PYRAF task {\tt multi-drizzle}, which
includes cosmic ray rejection and correction of geometric
distortion. Detection images are built by producing a surface brightness
profile of the galaxy using either the {\tt PYRAF} tasks ellipse and bmodel or
{\tt GALFIT} (Peng \etal 2002), as appropriate. The surface brightness profile
and best fit Sersic models contain interesting information on the presence and
luminosity of stellar nuclei (see Ferrarese \etal 2006b). The model galaxy is
then subtracted and object detection is performed using {\tt SExtractor} 
with a threshold of 10 connected pixels at a level of 10-\si.  Then, a catalog
from SDSS is produced using the SDSS-DR5 on-line search
tool\footnote{http://cas.sdss.org/dr5/en/tools/search/radial.asp} to retrieve
all objects within a radius of 5\arcmin\ of the galactic nucleus. A matching
program then determines the coordinate offsets between the two catalogs by
applying a Hough transform: first, objects in each catalog are uniquely
matched to their nearest neighbor in the opposite catalog and the offset
between each pair of objects is calculated; any remaining unmatched objects
are discarded. Next, statistics on the offsets in RA and Dec are gathered by
an initial pass through all object pairs. Finally, successive rounds of sigma
clipping are performed to discard outliers and spurious detections, leaving a
minimum of 10 pairs of matched objects from which the overall RA and Dec
offsets are calculated. Observed offsets range from $0\farcs 01 - 1\farcs
2$, with errors $0\farcs 01-0\farcs 07$. Offset values are confirmed by
manually comparing coordinates of stars and/or galactic nuclei in the ACS
images to the SDSS database. No rotation of the world coordinate system is
required, as the residual r.m.s. scatter is much smaller than the uncertainty
on the position of the X-ray sources, which is dominated by the Chandra PSF
and the small number of counts for faint nuclear sources.

Published measurements by the ACSVCS group (Ferrarese \etal 2006b) were used
to estimate the B-band luminosity and stellar mass of the host galaxies.
Synthetic Vega B-band magnitudes (hereafter B) were obtained from the total
(i.e. as obtained from model fitting), extinction-corrected $g_0$ and $z_0$
band AB magnitudes, using a broad range of stellar population models (Bruzual
\& Charlot 2003) to compute the transformation to first order in the color
term. We find the transformation to be:
\begin{equation}
{\rm B} = g_0 + 0.193 + 0.026 (g_0-z_0)
\end{equation}
Since the B band is close in wavelength to the $g_0$ band,
the transformation introduces only a minimal uncertainty of order
0.01--0.02 mag. The resulting B magnitudes listed in Table~\ref{tab:cxo} typically super-cede the
photographic B$_{\rm T}$ magnitudes (see C\^ot\'e et al. 2004, and references
therein) and, unless otherwise indicated, will be used throughout this series (although for
VCC1030 and VCC1535, the B$_{\rm T}$ magnitudes are retained since HST
photometry was not available). For all the objects with HST
photometry, stellar masses were estimated from the $g_0$ and $z_0$ band AB
model magnitudes using the recipe of Bell \etal (2003): 
\begin{equation}
{\rm log} ({\rm M}^{\star}/{\rm L}_{g_{0}}) = 0.698 (g_0-z_0) - 0.367
\end{equation}
This recipe -- and its use of the
HST photometry -- was found to be more robust than similar ones that
use the 2MASS K-band data listed in Ferrarese \etal (2006b), perhaps due to the
difficulty of measuring fluxes of the lowest mass galaxies, or with
matching the measurement apertures between different types of
observation.  For the two objects with no HST photometry, we use B$_{\rm T}$
and K-band magnitudes and the coefficients provided in Bell \etal 
(2001) to compute:
\begin{equation}
{\rm log} (\rm M^{\star}/\rm L_{\rm B}) = 0.591 ({\rm B_T-K}) - 1.743
\end{equation}
which in these cases gives stellar masses that sit well with
objects of comparable luminosity and measured with HST. As noted by
Bell et al, the mass to light ratios calculated with these recipes
have systematic uncertainties of some 0.2 dex arising from the
assumed initial mass function (a Salpeter function was used in the
derivation of the coefficients used here).
%
\section{Black hole masses}
\label{sec:bhm}

In order to construct the distribution of Eddington ratios for our
sample, we first need to estimate the masses of the (putative) SMBHs;
throughout this Section, we shall {\it assume} that a SMBH exists at
the center of every galaxy in the sample; this working hypothesis will
be tested and discussed in ~\S~\ref{sec:res}. 

Although bulge stellar velocity dispersion is arguably the best
estimator of \bhm\ (e.g. Bernardi \etal 2007), there is considerable
interest in comparing \si-based estimates with other estimates. For
the most luminous galaxies, such as brightest cluster galaxies,
$\sigma$ and optical luminosity (\lb) predict different \bhm\ (Lauer
et al. 2007a) pointing towards a break down of at least one of the two
scaling relations or to a departure from a simple power law.
Under the assumption that a SMBH exists at the center of each targeted
galaxy, in the following we compare the \bhm\ values obtained by
employing different empirical scaling relations, specifically the
mass-bulge luminosity (\bhm-\lb) or the mass-dispersion velocity
(\bhm-\si) relation. Unless otherwise indicated, we will employ the
scalings given by Ferrarese \& Ford (2005; FF05 hereafter).

\begin{figure}
\center{
\includegraphics[angle=0,scale=.43]{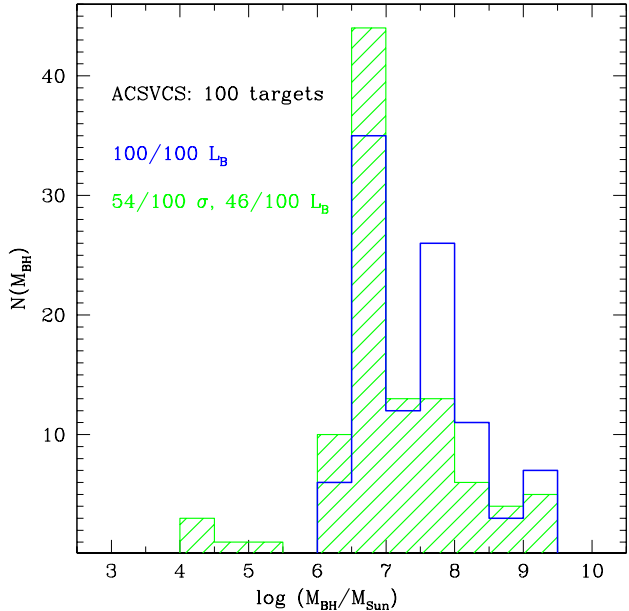}
\caption{Distribution of \bhm\ for the 100 spheroids belonging to the
ACS Virgo Cluster Survey sample, adopting different mass estimators based on
the properties of the host galaxies: \bhm-\lb\ for all galaxies (blue line),
and \bhm-$\sigma$ for the galaxies with `secure' (see \S~\ref{sec:bhm})
$\sigma$ and \bhm-\lb\ otherwise (green shaded). The latter is the fiducial
distribution adopted in this paper. Scalings are from Ferrarese \& Ford
(2005).
\label{fig:distr_bhmass}}}
\end{figure}
%
Velocity dispersions are available in the literature for 74 out
of the 100 targets belonging to the ACSVCS, from a variety
of sources. In this work we make use of a compilation kindly provided
by Lauren MacArthur (MacArthur et al. 2007).  However, as black hole 
mass is a steep function of stellar velocity dispersion, imprecise or
inaccurate spectroscopic measurements could introduce significant uncertainty and
bias to the black hole mass estimates. After considering different
velocity dispersion values from 12 different sources in literature, and investigating
the instrumental resolutions and S/N ratios of the original
measurements, a subset of high quality velocity dispersions was
identified, yielding a `secure' sub sample of 54 galaxies (see
MacArthur et al.\ 2007 for details).

Figure~\ref{fig:distr_bhmass} illustrates the \bhm\ distribution of
the entire ACSVCS sample as obtained using the two different mass
tracers: \lb-based masses (blue histogram) tend to be higher than
$\sigma$-based masses (green shaded histogram),
particularly at the low-luminosity and low-mass end. While this is a
known fact (e.g. Bernardi \etal 2007), in this specific case it can be
due to a combination of underestimated $\sigma$, overestimated bulge
luminosity \lb\ (because of low bulge fraction), as well as different
slopes of the \bhm-\lb\ and \bhm-$\sigma$ relations.  Irrespectively
of the chosen tracer, however, the distribution of the ACSVCS sample
peaks below 10$^7$ M$_{\odot}$, where very few direct \bhm\,
measurements are available.

\begin{figure}
\center{
\includegraphics[angle=-90,scale=.35]{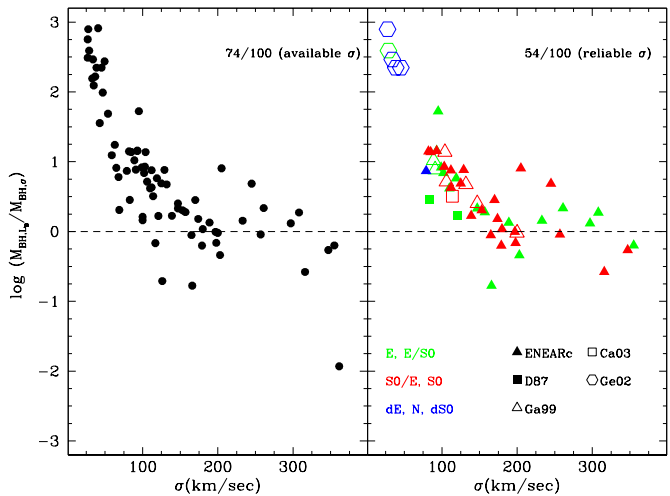}
\caption{{\it Left:} Ratio of \bhm\ as estimated from the \bhm-\si\
relation to that estimated from the \bhm-\lb\ relation (both scalings
are from FF05) for the 74 targets for which $\sigma$ measurements are
available in the literature. {\it Right:} Same as left, only for the
54 objects with 'secure' $\sigma$ measurements.  Here, targets are
color-coded for different morphological types, showing that the
discrepancy at the low-mass/low-luminosity end cannot be entirely due
to a reduced bulge fraction. References are: Bernardi \etal (2002; ENEARc);
Davis \etal (1987; D87); Gavazzi \etal (1999; Ga99); Caldwell \etal (2003;
Ca03) and Geha \etal (2002; Ge02). \label{fig:msig}}}
\end{figure}

Studies of active galaxies indicate that the \bhm-$\sigma$ relation
extends down to the masses probed by our sample (Barth, Greene \& Ho
2005), supporting our working hypothesis that $\sigma$ provides the
best estimate of \bhm.  Therefore, the results presented in the rest
of the paper will be based on this `secure' sample of stellar velocity
dispersions (for 54/100 targets), and on \lb\ for the remaining
targets.  This fiducial \bhm\ distribution adopted in this paper is
shown in Figure~\ref{fig:distr_bhmass} as a green shaded histogram; making use of \si\ values introduces a minor correction  at the low-mass end.
Black hole masses based on this secure sample are further compared to
\lb-based \bhm\ for different morphological types in the right panel of 
Figure~\ref{fig:msig}.  While a full discussion on the comparison
between different \bhm\ tracers is beyond the scope of this paper, the
plots show that -- although the mismatch is reduced when considering
only high quality $\sigma$ -- it is present even for pure ellipticals
and thus cannot be explained entirely with varying bulge fraction,
pointing instead towards a genuine breaking-down of least one of the
two scaling relations at the low mass end.
The conclusions of this paper are not significantly affected if different
scalings/black hole mass indicators are employed (such as Tremaine \etal 2002 or Marconi \& Hunt
2003, for \bhm-\si\ and \bhm-\lb, respectively).
\begin{figure*}[t!]
\center{
\center{\includegraphics[angle=0,scale=.99]{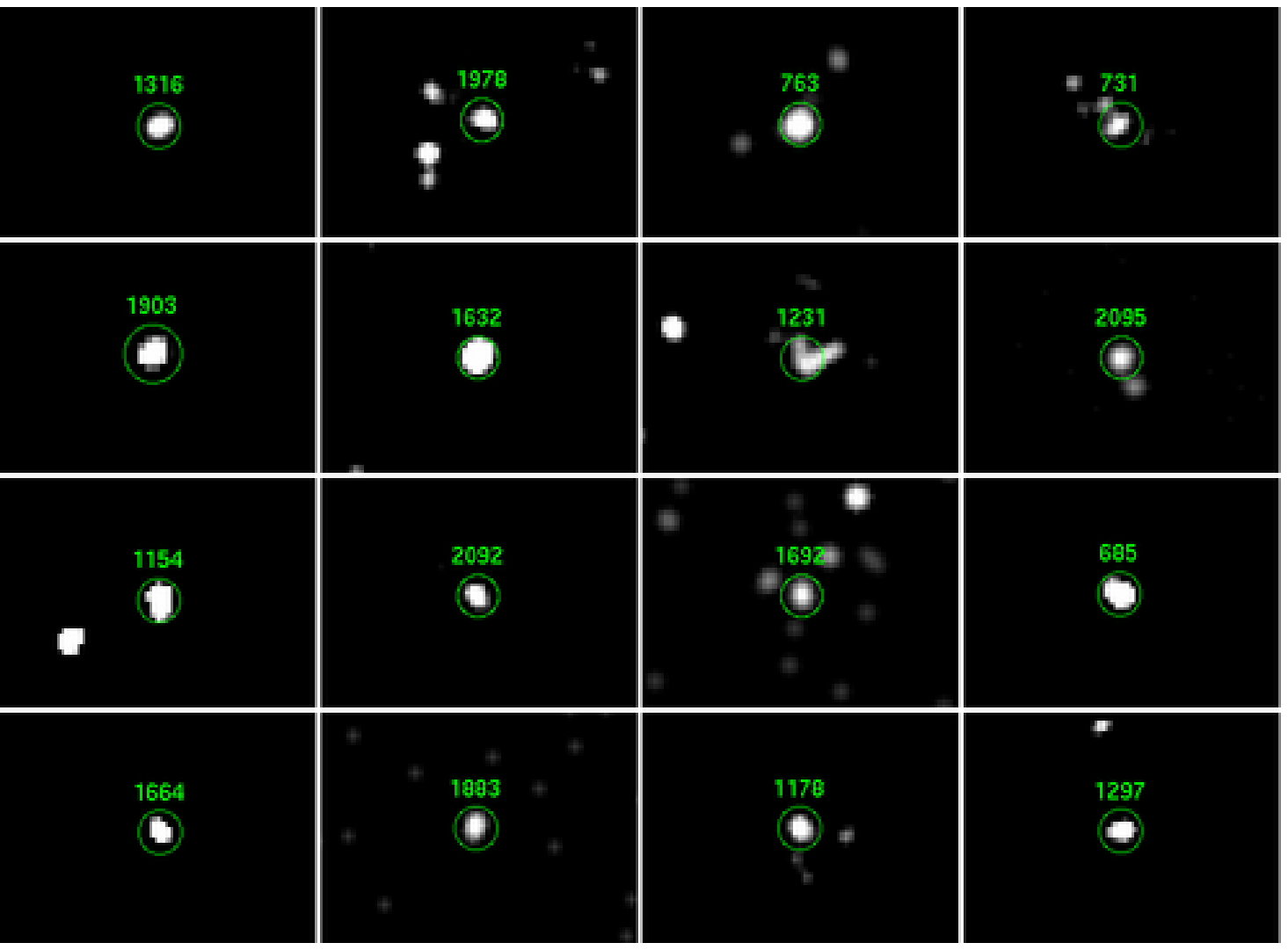}}
\caption{\cxo ACIS-S images of the 16 detected X-ray
nuclei, smoothed with a Gaussian of $\sigma$=3\arcsec. The green circle
represent the count extraction region, centered on the {\tt wavdetect}
centroid. Four of the detected targets belong to the large program survey and
were observed during Cycle 8 (VCC2092, VCC1692, VCC1883, VCC1178); three
targets have archival observations but no literature reference for the nuclear
source (VCC1903, VCC1231, VCC2095). The faintest detection, both in terms
of nuclear X-ray luminosity and host galaxy magnitude ($M_B$=$-$16.85), is 
VCC1297, and was previously reported by Soria \etal (2006b).
\label{fig:cxo}}}
\end{figure*}

\section{Results}
\label{sec:res}

Table~\ref{tab:cxo} lists the nuclear X-ray properties for the 32 targets
under analysis.  We detect point-like X-ray emission from a position
consistent with the optical nucleus in 16 targets, 4 of which (VCC2092,
VCC1692; VCC1883, VCC1178) belong to the list of new snapshot (5.4 ksec) \cxo
observations. A montage of the ACIS-S images of the detected 
nuclei is shown in Figure~\ref{fig:cxo}. For the 16 targets with archival data, we were able to compare
our results on the nuclear X-ray sources (or lack thereof) with the literature
in 11 cases, finding good agreement with the published values: after
re-scaling to the same distances, we obtain an average luminosity difference of
0.03 dex, with a scatter of 0.17 dex.
No fluxes/upper limits had been published for the nuclear emission for
the remaining 5 targets. We briefly comment on them below. VCC1903: the \cxo data for this galaxy have 
been analyzed and discussed in a number of publications, focusing
either on the diffuse X-ray emission properties or the X-ray binary
population. VCC881, VCC1535: the data for these galaxies have been
discussed in the context of X-ray binary population studies, excluding
the central regions (e.g. Sivakoff \etal 2007). VCC1231, VCC2095: no
publication has been found regarding these
\cxo datasets.

A fundamental point to be addressed is the nature of the detected nuclear
X-ray emission. In principle, hard X-rays provide us with some
clear-cut diagnostics for accretion-powered emission, as a result of
non-thermal processes such as Comptonization. 
In particular, here we ask the question whether
accretion-powered emission from a SMBH is at the
origin of the detected nuclear sources. In the following, we shall
carefully address the issue of contamination from background X-ray
sources as well as low mass X-ray binaries (\S~\ref{ssec:nature}),
which are the major source of concern. This is closely related to the
issue of whether SMBHs exist and/or they are {\it detectable} at the
center of faint spheroids which may host compact stellar clusters
(\S~\ref{ssec:small}). The Eddington distribution of the
detected nuclei is presented in \S~\ref{ssec:edd}, and discussed in the
context of the various model for inefficient accretion and mechanical
feedback from SMBHs.

\subsection{Origin of the detected X-ray emission}
\label{ssec:nature}

We argue that the detection of point-like X-ray emission from a position
coincident with the optical nucleus is unlikely to be due to any process other
than accretion onto a nuclear SMBH.  Based on the results by Rosati \etal
(2002), we estimate that the chance of detecting a background X-ray source
within the \cxo PSF at 1.5 keV (convolved with the positional uncertainty) is
lower than 10$^{-6}$.  Hence, the most likely contamination arises from
LMXBs. In a broad stellar mass range, {\it and in the absence of a nuclear
star cluster} (see~\S~\ref{ssec:small}), the total number of LMXBs and their
cumulative X-ray luminosity are proportional to the stellar mass of the host
galaxy, M$_{\star}$ (Gilfanov 2004; Kim \& Fabbiano 2004;
Humphrey \& Buote 2006).  The number $n_X$ of expected sources per
unit stellar mass above a certain luminosity threshold can be estimated from
the X-ray luminosity function for LMXBs (e.g. Gilfanov 2004).  In turn, the number of
expected sources within the \cxo PSF (convolved with the positional
uncertainty) is given by $n_X$ times M$_{\star, \rm PSF}$: the stellar mass
within the central aperture.  We estimated
M$_{\star, \rm PSF}$ for each galaxy from the archival ACS images, adopting
the same procedure as described in \S~\ref{sec:bhm}.  The number of expected
LMXBs above the X-ray luminosity of the detected nuclei turns out to be
typically lower than a few $10^{-2}$ for the most massive galaxies (which do
not harbor prominent stellar clusters at the center; Ferrarese \etal 2006b).
High mass X-ray binaries are not expected to contribute in early-type
galaxies, where star formation is nearly absent. As an example, the number of
expected LMXB with \lx$>\rm L_{\rm X,nucl}=4.7\times 10^{38}$ erg \se\
(VCC1178) is about 6 per $10^{11}$ \msun\ (this is obtained employing the
functional shape obtained by Gilfanov 2004 specifically for early-type galaxies). This means that fewer than 0.06
sources as bright/brighter than the detected nucleus are expected within the
central aperture, home to $\sim 9\times 10^8$ \msun. Given the shape of the
LMXB luminosity function at high luminosities (above a few $10^{38}$ erg \se),
we can also confidently rule out that the central X-ray source is due to a
collection of fainter LMXBs, as the integral $\int n_{X}\times {\rm L}_{\rm
X}~d{\rm L}_{\rm X}$ is dominated by the luminosity term. The same conclusion
is reached by Sivakoff \etal (2007) in an extensive study of the X-ray
luminosity function of globular clusters in early-type galaxies
(see~\S~\ref{ssec:small}).  As an example, in the case of VCC1178, the number
of expected nuclear LMXBs brighter then 1/10 of the detected nucleus is less
than 0.6.
However,
massive star clusters have been shown to become more and more common
at the center of spheroids moving down the mass function (Ferrarese
\etal 2006a), and may well increase the chance of harboring bright
X-ray binaries. This is further explored in the next Section.

\subsection{SMBHs in low-mass spheroids}
\label{ssec:small}

Possibly the most noteworthy result of this study is the detection of nuclear
hard X-ray emission from faint early-type galaxies: in particular, two of the
detected nuclei are hosted in galaxies with absolute B magnitudes lower than
$-$18: VCC1178(=N4464) and VCC1297(=N4486B) have $M_{\rm B}$=$-$17.68 and $M_{\rm B}$=$-$16.91,
respectively. 

From an observational standpoint, the very existence of SMBHs (of the
same sort that define the known scalings in massive galaxies) in faint
inactive galaxies remains questionable. Ferrarese \etal (2006a)
suggest that the creation of a `central massive object', SMBH or
compact stellar nucleus, would be the natural byproduct of galaxy
evolution, with the former being more common in massive bright
galaxies ($M_{\rm B}$ brighter than $-$20), and the latter dominating
--possibly taking over-- at magnitudes fainter than $-18$. 

This finds
support in semi-analytical models which follow the formation and
evolution of black holes seeds formed at high redshift in the context of
hierarchical cosmologies.
On one side, SMBH formation mechanisms seem to be more efficient in halos of 
high mass; on the other, low-mass objects are more likely to eject their
nuclear SMBH following a major merger as a result of gravitational recoil.
The combination of these two effects may lead to a lower black hole occupation
fraction in low mass galaxies at red-shift zero (Volonteri
\etal 2007a,b; it should be stressed however, that in this scenario 
nuclear SMBHs and compact star clusters are not necessarily mutually
exclusive). Observationally, the fraction of X-ray detectable SMBHs (assuming that they
can indeed be distinguished from bright LMXBs) would naturally place a lower
limit on the black hole occupation fraction in low-mass spheroids. We investigate this
below.

\begin{figure}
\hspace{0.9cm}\includegraphics[angle=0,width=6.5cm]{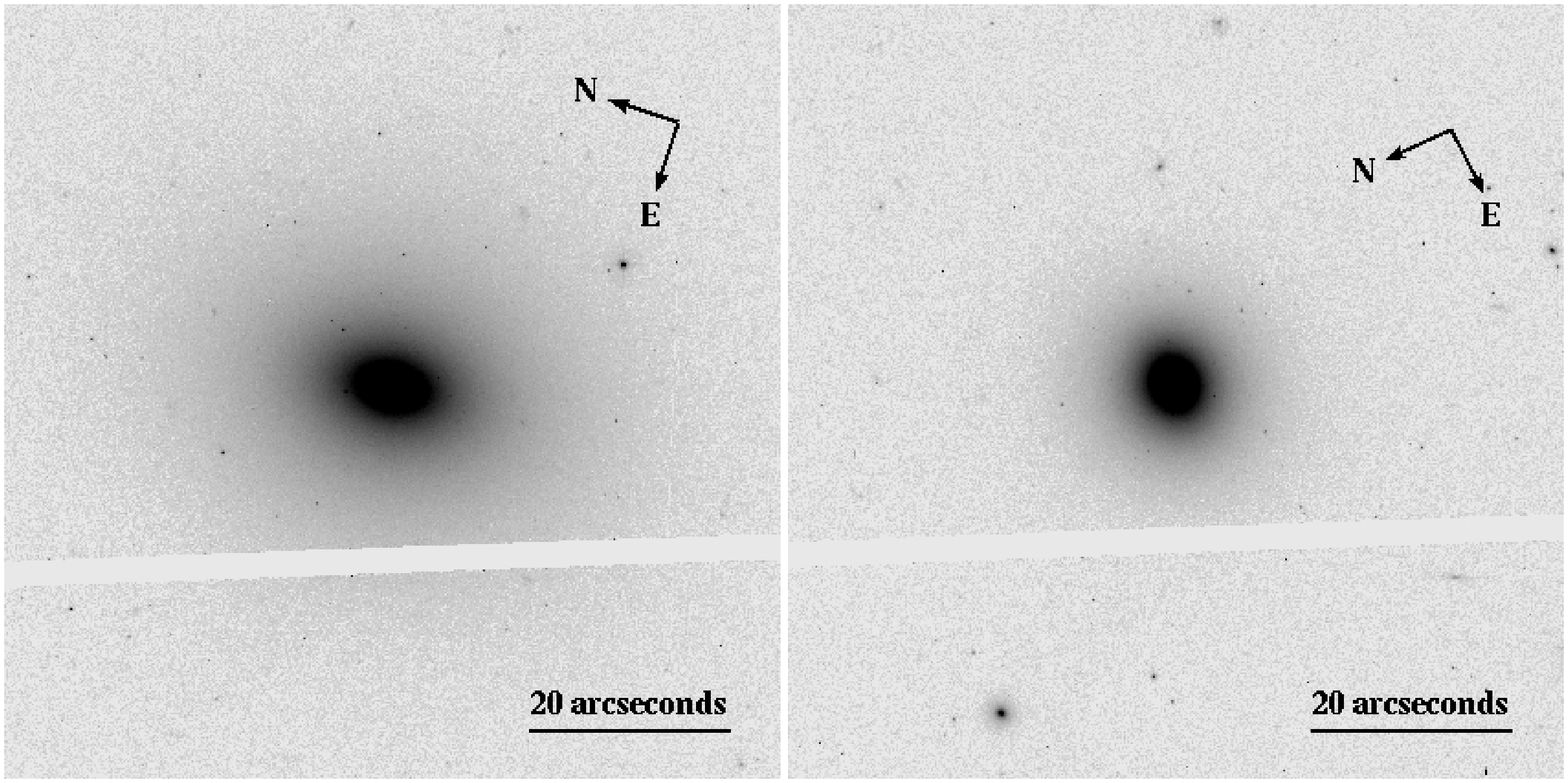}\\
\hspace{1.3cm}\includegraphics[angle=0,width=7.5cm]{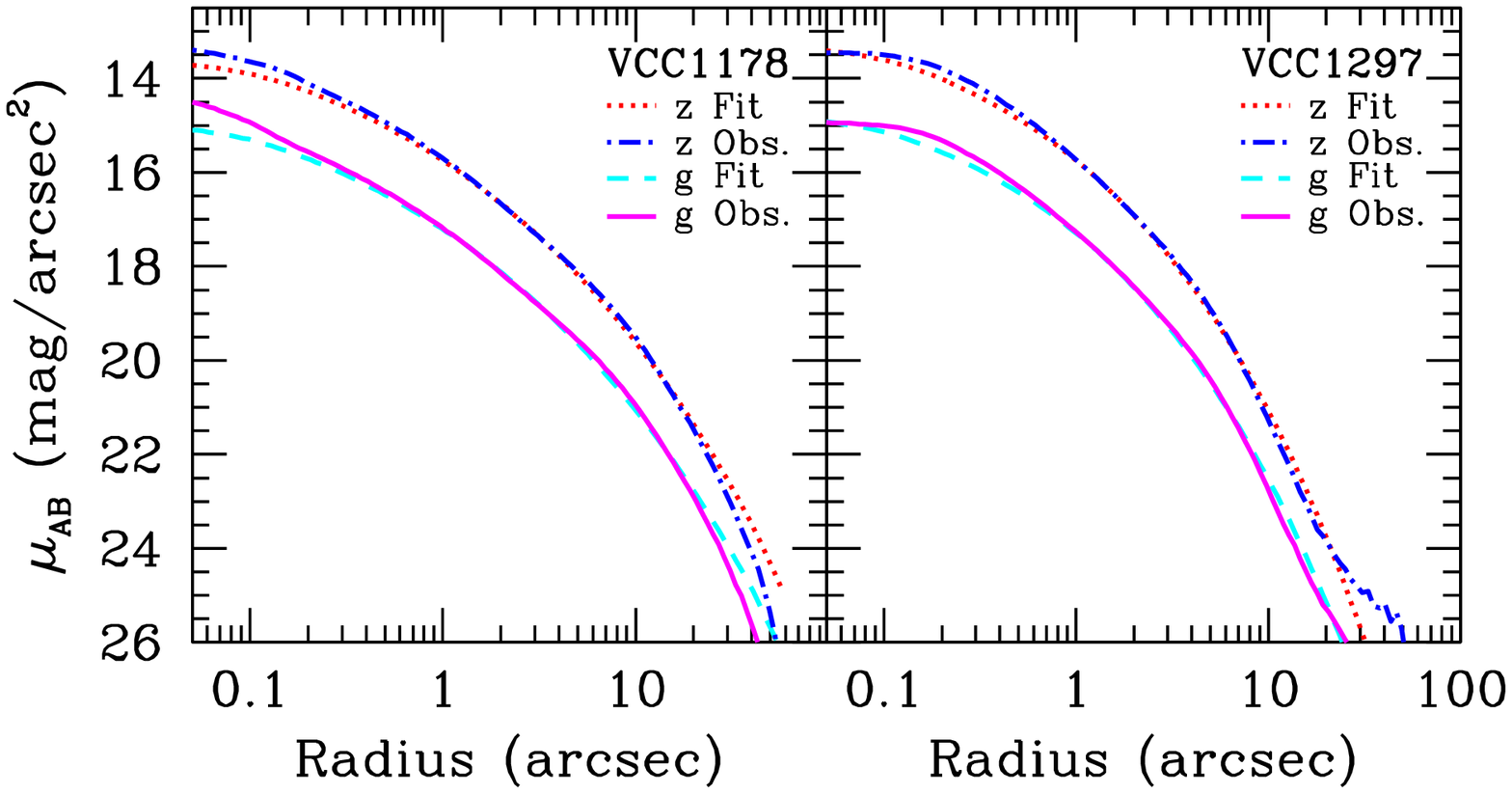}
\caption{{\it Upper panels:} $g'$-band HST images of VCC1178 (left) and VCC1297 (right),
as observed in the ACS Virgo Cluster Survey (C{\^o}t\'e\ et al. 2004).  The
field of view is approximately 72\arcsec\ on a side.  {\it Lower panels:}
azimuthally averaged $g'$ (solid magenta) and $z'$-band (dashed-dotted blue)
surface brightness profiles for VCC1178 and VCC1297 (the actual errorbars are
too small to be visible on the plots). Best-fit single-component Sersic models
(cyan and red) 
are over-plotted for each galaxy. A nuclear cusp is just visible in the inner
$\sim0\farcs2$ of the VCC1178 profile, in contrast to the flattened nuclear
profile observed in VCC1297.
\label{fig:small}}
\end{figure}

As shown in Figure~\ref{fig:small}, VCC1178 and VCC1297 do not have a
particularly prominent nuclear star cluster, consistent with the findings of
Ferrarese
\etal (2006b). 
However, the case for a SMBH is quite strong in VCC1297(=N4486B): based on
data from the Wide Field Planetary Camera 2 (WFPC2), Lauer et al. (1996)
showed evidence for a central double nucleus in this galaxy; subsequently,
based on stellar kinematics studies, Kormendy \etal (1997) derived a nuclear
`dark mass' of $6\times 10^8$ \msun, both arguing in favour of a nuclear SMBH.
A small excess with respect to the model fit in the inner region of the
profile is just noticeable in VCC1178. We note however that this conclusion is
highly dependent upon the assumed form of the profile of the underlying faint
galaxy, which has no fundamental reason to follow exactly a Sersic-law. In
fact, Lauer \etal (2007b) performed deconvolved HST ACS surface photometry
study of a sample of early-type galaxies, including VCC1178(=N4464), and found
evidence for a nuclear source in this system by modeling the profile with a
Nuker law. Assuming that the excess flux is due to a nuclear star cluster in
this galaxy, we estimate its luminosity to be approximately $2.8\pm 0.3\times
10^{7}$ L$_{\odot, z}$, where the error bar is the semi-difference of the
results obtained from two different methods that should bracket the true
answer: i) fitting a point spread function + Sersic model; ii) aperture
photometry on the residuals of the Sersic fit within a 0\arcsec.5 radius
aperture.  This corresponds to a stellar mass between $3.7-4.6\times 10^7$
\msun, where we have adopted a mass-to-light ratio $\Upsilon_z=1.45$, to
ensure a proper comparison with the work by Sivakoff
\etal (2007), which provides an expression for the expected number of bright
LMXBs specifically in globular clusters (rather than averaged over the entire
galaxy).
%

It is known that, while hosting a small percentage of the galaxy
stellar mass, globular clusters are home to about 50 per cent of the
observed LMXBs. In this environment, the number of expected LMXB
sources scales non-linearly with the cluster mass (Sivakoff \etal
2007); this also leads to the prediction that high X-ray luminosity
clusters (with super-Eddington luminosities for 3 \msun)  contain {\it a single LMXB}.
Sivakoff \etal (2007) derive an expression for the expected number
$n_X$ of LMXBs brighter than $3.2\times 10^{38}$ erg \se\ (the luminosity
limit is set by the sample completeness) in a star
cluster of stellar mass $M$, half mass radius $r_{\rm h,corr}$, and color
($g$-$z$):
\begin{equation}
n_X = 8\times 10^{-2} \big({M}/{10^6 M_{\odot}}\big)^{1.237} 10^{0.9(g-z)}
\big({r_{\rm h,corr}}/{1
{\rm pc}}\big)^{-2.22}
\end{equation}
where $r_{\rm h,corr}=r_{\rm h}\times 10^{0.17[(g-z)-1.2]}$.  An estimate of
the half mass radius of the star cluster in VCC1178 is obtained by measuring
the light in the residuals as a function of photometric aperture, and varies
between 25 and 30 pc. Together with the fitted luminosities, this implies
$n_X$=0.06--0.12 (obtained by adopting the fitted parameters with method i)
and ii), respectively).  The expected number of LMXBs in a globular cluster
can be converted to a probability $P_X$ that there is at least one LMXB
brighter than the adopted X-ray luminosity threshold assuming a Poisson distribution: $P_X\simlt0.11$ (95 per cent
confidence). 

We note that this value represents a conservative estimate of the
actual probability contamination, in that it is estimated for a LMXB X-ray
luminosity threshold lower than any of the detected nuclei in our sample.
In addition, LMXBs are found more often in globular clusters with smaller half-mass
radii: since there is no correlation between the half-mass radius and the mass
in globular clusters (e.g. Jord\'{a}n \etal 2005), this simply implies that
LMXBs are found more often in {\em denser} environments, with higher encounter
rates (Sivakoff \etal 2007). The inferred half-mass radius of the (possible)
nuclear star cluster in VCC1178 (20-30 pc) is much higher than the typical
radius estimated for standard globular clusters (few pc, with a median value
of 2.2 pc in the work by Sivakoff \etal).  From this, we conclude that a
bright LMXB is unlikely to be at the origin of the observed nuclear emission
in VCC1178, with a maximum chance contamination of 11 per cent.

\begin{figure}
\center{
\includegraphics[angle=0,scale=.47]{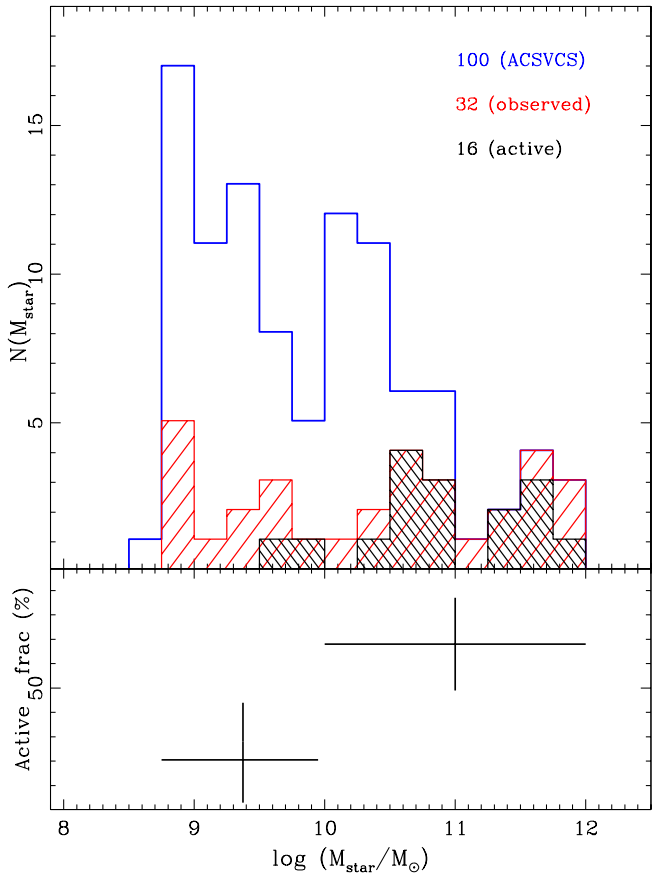}
\caption{{\it Upper panel:}
Stellar mass distribution of the early-type galaxies belonging to the
ACSVCS sample (blue histogram). The distribution of the 32 targets
considered in this work is shown as a red shaded histogram. The distribution
of the 16 targets with an X-ray active SMBH at the center is
represented by a black double-shaded histogram.  Galaxy stellar masses
M$_{\star}$ are estimated based on their B-band luminosity and colors,
as described in \S~\ref{ssec:HST}. Two of the detected nuclei
(VCC1178=N4464 and VCC1297=N4486B) belong to galaxies with stellar mass lower
than a few $10^{10}$ \msun, where the very survival (or existence) of
SMBH has been questioned (see Figure~\ref{fig:small} and discussion in
\S~\ref{ssec:small}).  {\it Lower panel:} incidence of X-ray nuclear activity as
a function of the host stellar mass. Only 2 out of the 12 galaxies with
stellar mass below $10^{10}$ \msun\ host an active SMBH, i.e. between 3--44
per cent. The percentage raises to 49--87 per cent above $10^{10}$ \msun,
where 14 out 20 galaxies host an active nuclear SMBH (numbers are given at the
95 per cent confidence level). \label{fig:mgal}}}
\end{figure}

The incidence of X-ray `active' (hereafter defined as detected in the X-ray band down to
our luminosity threshold of $\sim4\times 10^{38}$ erg \se) SMBHs as a function of
the host galaxy stellar mass, M$_{\star}$, is illustrated in the upper panel
of Figure~\ref{fig:mgal}. Splitting the sample in two mass bins, above and
below a stellar mass threshold of $10^{10}$ \msun, and by making use of
binomial statistics applied to small number of observed events (Gehrels 1986)
we are able to conclude that {\it the incidence of nuclear X-ray super-massive
black hole activity --down to our completeness limit of $\sim 4\times 10^{38}$
erg
\se-- increases with the stellar mass of the host} (see Figure~\ref{fig:mgal},
lower panel). Specifically: between 3 and 44 per cent of the galaxies with
stellar masses $<10^{10}$ \msun\ are found to host an active SMBH (2 out
12). The incidence of nuclear activity increases to between 49 and 87 per cent
in galaxies with stellar masses above $10^{10}$ \msun\ (14 out of 20 are
active; percentages are given at the 95 per cent confidence level).  For
comparison, in a recent comprehensive optical spectroscopic census of nuclear
activity associated with {\it late}-type galaxies in Virgo, Decarli \etal
(2007) find no AGN in galaxies with dynamical mass lower than $10^{10}$ \msun\
(in that work, line ratios are adopted in order to classify/distinguish AGN
from transition objects and/or HII regions, specifically: N$_{\rm
II}$/H$_{\alpha}>$ 0.6 unambiguously identifies AGN).

\subsection{Eddington-ratio distribution and nuclear SMBH feedback}
\label{ssec:edd}
\begin{figure}
\center{
\includegraphics[angle=0,scale=.43]{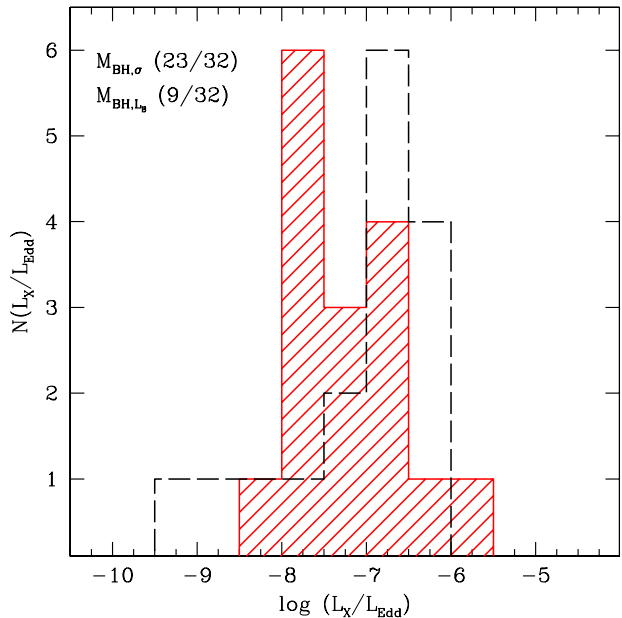}
\caption{\lx/\ledd\ (0.3-10 keV) distribution of the 32 targets under analysis. Black hole 
masses are calculated from dispersion velocities \si\ when deemed
'secure' (for 23/32 targets), from \lb\ otherwise.  The red histogram
represents the 16 detections, the dotted line represents luminosity
upper limits. \lx\ is measured between 0.3--10 keV. {\it Bolometric}
luminosities are likely a factor 8-60 higher (based on Marconi \etal
2004).
\label{fig:ledd}}}
\end{figure}

Having shown that point-like nuclear X-ray emission is likely due to
accretion onto a SMBH -- and under the assumption that the sample
galaxies all host a SMBH whose mass obeys the known scaling relations
defined by SMBHs in massive bright galaxies -- we can construct the
\lx/\ledd distribution of our sample, shown in Figure~\ref{fig:ledd}
by adopting the fiducial black hole mass distribution described in
\S~\ref{sec:bhm}.

For any plausible value of the bolometric correction $f_{\rm bol}=\rm L_{\rm
bol} $/\lx\ (which may vary between $\sim$8 and $\sim$60; Marconi \etal 2004),
the detected nuclei are highly sub-Eddington.  Under the conservative
assumption that only as little as 2$\%$ of accretion-driven emission is
emitted in the X-ray band, {the inferred L$_{\rm bol}$/\ledd\ ratios do not
exceed $10^{-4.7}$ for this sub-sample (inferred for VCC1883, the highest
\lx/\ledd nucleus among our 16 snapshot observations)}. Similarly, the upper
limit to the {\it average} X-ray luminosity in the stacked image of the 12
undetected-nuclei with snapshot observations (Figure~\ref{fig:stack}, left
panel) amounts to $3.8 \times 10^{37}$ erg \se, or $\langle$\lx/\ledd$\rangle
\simlt3\times 10^{-8}$, over 0.3-10 keV) for an average black hole mass of
$9.3\times 10^6$ \msun.
Similar results are obtained by Santra \etal (2007) for a sample of 13
early-type galaxies in the core of the Perseus cluster with a deep \cxo
exposure, as well as from Soria \etal (2006a) and Pellegrini (2005). 
Following their approach (see eqs. 4 and 5 in Soria \etal 2006a), we can compare the
measured X-ray luminosities to the bolometric accretion power L$_{\rm acc,
IS}$ expected from Bondi accretion of the interstellar medium: L$_{\rm acc,
IS}=\eta\dot{M}_{\rm B}c^2$, where the radiative efficiency $\eta$ is a
fraction $f_r$ of the total accretion efficiency $\eta'$, and $\dot{M}_{\rm B}$ -- the Bondi 
accretion rate -- can be expressed as:
\begin{eqnarray}
\dot{M}_{\rm B} &=& 1.6 \times 10^{-5} 
	\left(\frac{M_{\rm BH}}{10^8 M_{\odot}}\right)^2\, 
	\left(\frac{0.5 {\rm ~keV}}{kT}\right)^{3/2}\nonumber\\
	&&\times \left(\frac{n_0}{0.01 {\rm ~cm}^{-3}}\right) \ \
	M_{\odot}\ {\rm yr}^{-1},
\end{eqnarray}
being $T$ and $n_0$ the temperature and density of the hot interstellar gas
($k$ is the Boltzmann constant).
Adopting conservative values of 0.01 cm$^{-3}$ for $n_0$ (in
order to minimize L$_{\rm acc,IS}$) and for the range of temperatures which we
infer for the hot gas (\S~3.2), we obtain:

\begin{equation}
\frac{L_{\rm{X}}}{0.1\dot{M}_{\rm{B}}c^2} 
	= f_{\rm X} f_{\rm r} \left(\frac{\eta'}{0.1}\right) 
	\left(\frac{\dot{m}}{\dot{m}_{\rm B}}\right) 
	\sim 2 \times 10^{-5}{\rm \,-\,}0.6,
\end{equation} 

where $f_{\rm X}=1/f_{\rm bol}$ ($0.02\simlt f_{\rm X}\simlt 0.12$; Marconi \etal 2004),
$\dot{m}$ and $\dot{m}_{\rm B}$ are Eddington-scaled $\dot {M}$ and
$\dot{M}_{\rm B}$, respectively, and $c$ is the speed of light.
All the detected nuclei in the AMUSE-Virgo
sub-sample presented here are under-luminous with respect to Bondi accretion
from the interstellar medium.

The accretion mode
responsible for powering low-luminosity black holes is still a matter of
debate.  Observations of highly sub-Eddington black holes, most notably the
Galactic Center Sag A$^{\star}$, paved the way to radiatively inefficient
accretion flow models (RIAFs). Advection-dominated accretion flows (ADAFs;
Ichimaru 1977; Narayan \& Yi 1994) are popular analytical models for the
dynamics of RIAFs at low accretion rates. However, they face a number of
difficulties. In particular, Blandford \& Begelman (1999; 2004) argued that
the accreting gas in an ADAF is generically unbound and free to escape to
infinity, and elaborated an alternative model, named the adiabatic inflow
outflow solution (ADIOS). Here the key notion is that the excess energy and
angular momentum is lost to a wind at all radii; the final accretion rate into
the hole may be only a tiny fraction of the mass supply at large radii. This
is a generalization of and an alternative to an advective inflow.  Deep \cxo
observations of nearby massive ellipticals (Allen \etal 2006), have shown that
a tight, almost linear, correlation exists between the Bondi accretion rate
and the jet kinetic power (measured from the $p\times dV$ work exerted on
X-ray cavities). The correlation implies that a substantial fraction of the
energy associated with gas entering the Bondi radius must be dissipated
mechanically, via jets/outflows.

This is closely related to the role of SMBH feedback in galaxy
evolution (see e.g. McNamara \& Nulsen 2007). Semi-analytical models applied to dark matter simulations
for the growth and evolution of cosmic structures have recently
emphasized the role of {\it mechanical}, rather than radiated, SMBH
feedback. A prolonged phase of low-level accretion, resulting in sub-Eddington
luminosities, proves to be effective at quenching star
formation.  It is worth stressing
though, that in this formulation SMBH feedback plays a role in the
most massive bright galaxies only: once star formation has halted,
these massive red galaxies continue to grow through merging. This
allows the brightest cluster galaxies to gain a factor of 2 or 3 in
mass without significant star formation.  However, it is not clear whether 
such a mechanism might switch off at or be still important at low-masses. 
%
Merloni \& Heinz (2007) have addressed this issue of mechanical SMBH feedback
by comparing a sample of 15 sub-Eddington nuclei for which information on the
Bondi rate, kinetic power, and radiative power is available.  For these
objects, they find that the Eddington-scaled black hole kinetic power, $\rm
L_{\rm kin}$ (which is a proxy for $\rm L_{\rm BH}$ -- the SMBH power feedback
-- in the formalism of
Croton \etal 2006) scales with the nuclear X-ray luminosity (2-10 keV) according to the
following relation: 
\begin{equation} {\rm log}(\rm L_{\rm kin}/\rm L_{\rm
Edd})=0.49 {\rm log}\lambda_{\rm X}-0.78
\end{equation} 
where $\lambda_{\rm X}=5\times {\rm L}_{\rm (2-10~keV)}/$\ledd\
(with a scatter of 0.39 dex).

Applying the Merloni \& Heinz scaling to our sample (where with are
arbitrarily assuming that the same correlation applies to lower-mass,
lower-luminosity objects
spheroids\footnote{In fact 4 out of 32 objects in our sample belong to
the sample examined by Merloni \& Heinz.}), the measured X-ray
luminosities translate into Eddington-scaled kinetic luminosity in the
range $\sim 10^{-3}$ (VCC1664) to $\sim 10^{-5}$ (VCC1978), which suggest that
energy feedback might be effective even in low mass spheroids.

\section{Summary and conclusions}
\label{sec:sum}

This paper presents the first \cxo results of AMUSE-Virgo, a
multi-wavelength survey of early type galaxies in the Virgo cluster,
aimed at investigating the incidence and activity of super-massive
black holes in the nuclei of 100 local nearby spheroids. The ACSVCS
sample (C{\^o}t\'e\ etal 2004) is selected based on the properties of the host
galaxies, and therefore it is an unbiased census of nuclear activity
as a function of host galaxy stellar mass (and hence presumably black
hole mass). Since the stellar mass distribution of the galaxies peaks
below $10^{10}$ \msun, AMUSE-Virgo will provide us the deepest census
of low-level accretion-powered activity in early type galaxies over an
unprecedented range of masses.  In this paper, we combine \cxo results from the first
16 targets with the analysis of archival data of 16, typically more
massive, targets. The absolute B magnitudes of this sample of 32
objects range from $M_B=-22.5$ to $M_B=-15.0$.

The main results of this study can be summarized as follows:

\begin{itemize}

\item We detect point-like X-ray emission from a position consistent with
the optical nucleus in 50 per cent of the targets. 12 detection out of 16 belong to
the archival observations, but only 9 out of those 12 were previously reported
in the literature. The remaining 4 detections (VCC2092, VCC1692, VCC1833,
VCC1178) were made using new snapshot observations of low-mass targets.   

\item Two of the detected nuclei (VCC1178 and VCC1297, having
\lx=$4.7\times 10^{38}$ and $2.6\times 10^{38}$ erg \se, respectively)
are hosted in galaxies with absolute B magnitude fainter than $-18$ ($M_B =
-17.68$ and $-$16.91), or host galaxy stellar mass lower than $10^{10}$ \msun\
(M$_{\star} = 8.1\times 10^{9}$ \msun\ and $5.1\times 10^9$ \msun). At these
luminosities, massive stellar clusters are known to become increasingly
common, and have been suggested to possibly even replace SMBHs (of the kind
that define empirical scaling relations at the bright end; Ferrarese \etal
2006a).

{\item Analysis of archival ACS HST images reveals a slight excess
in the surface brightness profile of VCC1178, with respect to a
Sersic model. We conservatively interpret this as due to a nuclear
star cluster. The inferred stellar mass does not exceed a few $10^7$
\msun, implying a less than 11 per cent probability that the nuclear
X-ray source is a solar mass compact object based on results by
Sivakoff et al. (2007).}

\item 
After carefully addressing possible contamination from low mass X-ray
binaries in the remaining objects (based on the luminosity function by
Gilfanov 2004), we conclude that {\it the nuclear X-ray sources are
most likely due to low-level accretion-powered activity from a super-massive
black hole}.

\item Between 3 and 44 per cent (95 per cent confidence level) of the galaxies with stellar masses lower than $10^{10}$
\msun\  harbor an X-ray active SMBH -- down to our completeness limit of $\sim
4\times 10^{38}$ erg \se. The fraction of galaxies hosting an active SMBH
increases to between 49 and 87 per cent for host masses above $10^{10}$. Even
with only a third of the sample (32/100 galaxies), this study shows that there
is {\it statistically significant increase in the incidence of nuclear
activity towards the high mass end}, consistent with what found in late-type
Virgo galaxies (Decarli \etal 2007). This should be folded with the actual
black hole mass function in order to properly constrain the distribution of
nuclear activity.

\item The upper limit to the {\it average} X-ray luminosity in the
stacked image of the 12 undetected-nuclei with snapshot (5.4 ks) \cxo observations
amounts to $3.8 \times 10^{37}$ erg \se, or $\langle$L$_{\rm (0.3-10~keV)}$/\ledd$\rangle
<3\times 10^{-8}$ for an average black hole mass of $9.3\times 10^6$ \msun.

\item Based on `fiducial' values for the central black hole mass (based on
'secure' measurements of the dispersion velocity for 24 targets, and on B
magnitude otherwise) the ratio L$_{\rm (0.3-10~keV)}$/\ledd varies between
$10^{-8.4}$ and $10^{-5.9}$ for the detected nuclei. The detected nuclei are
under-luminous with respect to Bondi accretion from the interstellar gas. In
agreement with earlier works (e.g. Pellegrini 2005; Soria \etal 2006a,b;
Santra \etal 2007), this argues for an inefficient accretion mechanism, albeit
our results can not break the degeneracy between intrinsically low radiative
efficiency and/or drastically reduced mass accretion rate onto the black hole
(owing to outflows/winds).

\end{itemize}

A crucial question still to be addressed is that of the amount of power
released in the form of kinetic energy. According to a recent study by Merloni
\& Heinz (2007), a (non-linear) correlation exists between the
Eddington-scaled kinetic power and the bolometric luminosity. The
non-linearity implies that the relative amount of power dissipated by these
nuclei in the form of mechanical power decreases toward low X-ray
luminosities. If the same scaling is applied to our sub-sample of 16 detected
X-ray nuclei, the inferred kinetic power are between $\sim
10^{-5}-10^{-3}$\ledd, indicating that low-level SMBH feedback can be
effective in faint spheroids as well as in bright massive elliptical
galaxies.

Most of the galaxies yet to be observed as part of AMUSE-Virgo have stellar
masses around $10^{9.5}$ \msun, a mass range that remains largely unexplored
as far as low-level nuclear SMBH activity is concerned. \cxo observations of
68 additional faint galaxies are under way, and will further constrain the
fraction of galaxies that harbor a nuclear X-ray source. As shown by Ferrarese
\etal (2006a), massive nuclear star clusters become increasingly common down
the galaxy mass function, thereby increasing the chance of bright LMXB
contamination. Detection of a high
brightness temperature compact radio counterpart to the detected X-ray nuclei
would provide definitive evidence for an accreting SMBH, as no Galactic X-ray
binary can be possibly detected in the radio band at the Virgo cluster
distance with current instrumentation. In fact, with the exception of the
known radio sources VCC1226(M49), VCC1316(M87), VCC1978(M60), VCC763(M84),
VCC1535(N4526), VCC1632(M89), none of the sample galaxies have a detected
radio core brighter than the limiting flux density of 1.8 mJy at 1.4
GHz.\footnote{Data based on GOLDMine: http://goldmine.mib.infn.it/} At the
average distance of 16.5 Mpc, this corresponds to an upper limit to the radio
luminosity of $\nu \rm L_{\nu}< 8.3
\times 10^{38}$ erg \se.  
Deep radio observations of the targeted nuclei, with
the Very Large Array, are in progress, and will put further constraints on the
spectral energy distribution at low frequencies.
At the same time, while LMXBs are known to emit the bulk of the dissipated
accretion power in the X-ray band, SMBHs typically emit at longer wavelengths,
yielding bolometric corrections as high as 80 (Marconi \etal 2004); upcoming
mid-IR observations, with the \spi {\it Space Telescope}, will hopefully enable
us to estimate the bolometric luminosity of the detected nuclei, and to
uncover obscured SMBH activity.  

\acknowledgments E.G. is supported through \cxo Postdoctoral
 Fellowship grant number PF5-60037, issued by the \cxo X-Ray Center,
 which is operated by the Smithsonian Astrophysical Observatory for
 NASA under contract NAS8-03060. ~T.T.  acknowledges support from the
 NSF through CAREER award NSF-0642621, by the Sloan Foundation through
 a Sloan Research Fellowship, and by the Packard Foundation through a
 Packard Fellowship.~P.J.M. is supported by the TABASGO foundation in
 the form of a research fellowship. Support for this work was provided
 by NASA through Chandra Award Number 08900784 issued by the Chandra
 X-ray Observatory Center. This work is based on data obtained with
 the Chandra X-ray Observatory and archival data from the Hubble Space
 Telescope, obtained from the data archive at the Space Telescope
 Institute, 
which is operated by the association of Universities for Research in
Astronomy, Inc. for NASA under contract NAS5-26555.  We are grateful to
Lauren MacArthur for sharing with us her compilation of velocity
dispersion for Virgo galaxies. We acknowledge stimulating
conversations with Luca Ciotti, Darren Croton, Andr\'es Jord\'an, Andrea Merloni, Gregory Sivakoff and Marta Volonteri. We thank Andy Fabian and Tod
Lauer for useful comments and for pointing out relevant references, and the
referee, Silvia Pellegrini, for a detailed and constructive report.


\begin{thebibliography}{}

\bibitem[{{Abraham} {et~al.}(2007){Abraham}, {Nair}, {McCarthy}, {Glazebrook},
  {Mentuch}, {Yan}, {Savaglio}, {Crampton}, {Murowinski}, {Juneau}, {Le
  Borgne}, {Carlberg}, {J{\o}rgensen}, {Roth}, {Chen}, \& {Marzke}}]{Abr++07}
{Abraham}, R.~G., {Nair}, P., {McCarthy}, P.~J., {Glazebrook}, K., {Mentuch},
  E., {Yan}, H., {Savaglio}, S., {Crampton}, D., {Murowinski}, R., {Juneau},
  S., {Le Borgne}, D., {Carlberg}, R.~G., {J{\o}rgensen}, I., {Roth}, K.,
  {Chen}, H.-W., \& {Marzke}, R.~O. 2007, \apj, 669, 184

\bibitem[{{Allen} {et~al.}(2006){Allen}, {Dunn}, {Fabian}, {Taylor}, \&
  {Reynolds}}]{All++06}
{Allen}, S.~W., {Dunn}, R.~J.~H., {Fabian}, A.~C., {Taylor}, G.~B., \&
  {Reynolds}, C.~S. 2006, \mnras, 372, 21

\bibitem[{{Allen} {et~al.}(2000){Allen}, {Di Matteo}, \& {Fabian}}]{ADF00}
{Allen}, S.~W., {Di Matteo}, T., \& {Fabian}, A.~C. 2000, \mnras, 311, 493

\bibitem[{{Arnaud}(1996)}]{Arn96}
{Arnaud}, K.~A. 1996, 101, 17

\bibitem[{{Baganoff} {et~al.}(2003){Baganoff}, {Maeda}, {Morris}, {Bautz},
  {Brandt}, {Cui}, {Doty}, {Feigelson}, {Garmire}, {Pravdo}, {Ricker}, \&
  {Townsley}}]{Bag++03}
{Baganoff}, F.~K., {Maeda}, Y., {Morris}, M., {Bautz}, M.~W., {Brandt}, W.~N.,  {Cui}, W., {Doty}, J.~P., {Feigelson}, E.~D., {Garmire}, G.~P., {Pravdo},
  S.~H., {Ricker}, G.~R., \& {Townsley}, L.~K. 2003, \apj, 591, 891

\bibitem[{{Barth} {et~al.}(2005){Barth}, {Greene}, \& {Ho}}]{BGH05}
{Barth}, A.~J., {Greene}, J.~E., \& {Ho}, L.~C. 2005, \apjl, 619, L151

\bibitem[{{Barth} {et~al.}(2004)}]{Barth04}
{Barth}, A.~J., {Ho}, L.~C., Rutledge, R. E., Sargent, W. L. W. 2004, \apj, 607, 90


\bibitem[{{Begelman} {et~al.}(2006{\natexlab{b}}){Begelman}, {Volonteri}, \&
  {Rees}}]{BVR06}
{Begelman}, M.~C., {Volonteri}, M., \& {Rees}, M.~J. 2006{\natexlab{b}},
  \mnras, 370, 289

\bibitem[{{Bell} {et~al.}(2003){Bell}, {McIntosh}, {Katz}, \&
  {Weinberg}}]{Bel++03}
{Bell}, E.~F., {McIntosh}, D.~H., {Katz}, N., \& {Weinberg}, M.~D. 2003, \apjs,
  149, 289

\bibitem[{{Bell} \& {de Jong}(2001)}]{B+d01}
{Bell}, E.~F. \& {de Jong}, R.~S. 2001, \apj, 550, 212


\bibitem[{{Bernardi} {et~al.}(2007){Bernardi}, {Sheth}, {Tundo}, \&
  {Hyde}}]{Ber++07}
{Bernardi}, M., {Sheth}, R.~K., {Tundo}, E., \& {Hyde}, J.~B. 2007, \apj, 660,
  267

\bibitem[{{Bernardi} {et~al.}(2002){Bernardi}, {Alonso}, {da Costa}, {Willmer},
  {Wegner}, {Pellegrini}, {Rit{\'e}}, \& {Maia}}]{Ber++02}
{Bernardi}, M., {Alonso}, M.~V., {da Costa}, L.~N., {Willmer}, C.~N.~A.,
  {Wegner}, G., {Pellegrini}, P.~S., {Rit{\'e}}, C., \& {Maia}, M.~A.~G. 2002,
  \aj, 123, 2990

\bibitem[{{Biller} {et~al.}(2004){Biller}, {Jones}, {Forman}, {Kraft}, \&
  {Ensslin}}]{Bil++04}
{Biller}, B.~A., {Jones}, C., {Forman}, W.~R., {Kraft}, R., \& {Ensslin}, T.
  2004, \apj, 613, 238


\bibitem[{{Blandford} \& {Begelman}(2004)}]{B+B04}
{Blandford}, R.~D. \& {Begelman}, M.~C. 2004, \mnras, 349, 68


\bibitem[{{Blandford} \& {Begelman}(1999)}]{B+B99}
{Blandford}, R.~D. \& {Begelman}, M.~C. 1999, \mnras, 303, L1
%


\bibitem[{{Bundy} {et~al.}(2005{\natexlab{a}}){Bundy}, {Ellis}, \&
  {Conselice}}]{BEC05}
{Bundy}, K., {Ellis}, R.~S., \& {Conselice}, C.~J. 2005{\natexlab{a}}, \apj,
  625, 621

\bibitem[{{Bruzual} \& {Charlot}(2003)}]{B+C03}
{Bruzual}, G. \& {Charlot}, S. 2003, \mnras, 344, 1000

\bibitem[{{Caldwell} {et~al.}(2003{\natexlab{b}}){Caldwell}, {Rose}, \&
  {Concannon}}]{CRC03}
{Caldwell}, N., {Rose}, J.~A., \& {Concannon}, K.~D. 2003{\natexlab{b}}, \aj,
  125, 2891


\bibitem[{{Ciotti} \& {Ostriker}(2007)}]{C+O07}
{Ciotti}, L. \& {Ostriker}, J.~P. 2007, \apj, 665, 1038

\bibitem[{{Ciotti} {et~al.}(1991){Ciotti}, {Pellegrini}, {Renzini}, \&
  {D'Ercole}}]{Cio++91}
{Ciotti}, L., {Pellegrini}, S., {Renzini}, A., \& {D'Ercole}, A. 1991, \apj,
  376, 380

\bibitem[{{Churazov} {et~al.}(2005){Churazov}, {Sazonov}, {Sunyaev}, {Forman},
  {Jones}, \& {B{\"o}hringer}}]{Chu++05}
{Churazov}, E., {Sazonov}, S., {Sunyaev}, R., {Forman}, W., {Jones}, C., \&
  {B{\"o}hringer}, H. 2005, \mnras, 363, L91


\bibitem[{{C{\^o}t{\'e}} {et~al.}(2006){C{\^o}t{\'e}}, {Piatek}, {Ferrarese},
  {Jord{\'a}n}, {Merritt}, {Peng}, {Ha{\c s}egan}, {Blakeslee}, {Mei}, {West},  {Milosavljevi{\'c}}, \& {Tonry}}]{C++06}
{C{\^o}t{\'e}}, P., {Piatek}, S., {Ferrarese}, L., {Jord{\'a}n}, A., {Merritt},
  D., {Peng}, E.~W., {Ha{\c s}egan}, M., {Blakeslee}, J.~P., {Mei}, S., {West},
  M.~J., {Milosavljevi{\'c}}, M., \& {Tonry}, J.~L. 2006, \apjs, 165, 57

\bibitem[{{C{\^o}t{\'e}} {et~al.}(2004){C{\^o}t{\'e}}, {Blakeslee},
  {Ferrarese}, {Jord{\'a}n}, {Mei}, {Merritt}, {Milosavljevi{\'c}}, {Peng},
  {Tonry}, \& {West}}]{C++04}
{C{\^o}t{\'e}}, P., {Blakeslee}, J.~P., {Ferrarese}, L., {Jord{\'a}n}, A.,
  {Mei}, S., {Merritt}, D., {Milosavljevi{\'c}}, M., {Peng}, E.~W., {Tonry},
  J.~L., \& {West}, M.~J. 2004, \apjs, 153, 223


\bibitem[{{Cowie} {et~al.}(1996){Cowie}, {Songaila}, {Hu}, \&
  {Cohen}}]{Cow++96}
{Cowie}, L.~L., {Songaila}, A., {Hu}, E.~M., \& {Cohen}, J.~G. 1996, \aj, 112,
  839

\bibitem[{{Croton} {et~al.}(2006){Croton}, {Springel}, {White}, {De Lucia},
  {Frenk}, {Gao}, {Jenkins}, {Kauffmann}, {Navarro}, \& {Yoshida}}]{Cro++06}
{Croton}, D.~J., {Springel}, V., {White}, S.~D.~M., {De Lucia}, G., {Frenk},
  C.~S., {Gao}, L., {Jenkins}, A., {Kauffmann}, G., {Navarro}, J.~F., \&
  {Yoshida}, N. 2006, \mnras, 367, 864

\bibitem[{{Decarli} {et~al.}(2007){Decarli}, {Gavazzi}, {Arosio}, {Cortese},
  {Boselli}, {Bonfanti}, \& {Colpi}}]{Dec++07}
{Decarli}, R., {Gavazzi}, G., {Arosio}, I., {Cortese}, L., {Boselli}, A.,
  {Bonfanti}, C., \& {Colpi}, M. 2007, \mnras, 381, 136

\bibitem[{{Dalla Vecchia} {et~al.}(2004){Dalla Vecchia}, {Bower}, {Theuns},
  {Balogh}, {Mazzotta}, \& {Frenk}}]{Dal++04}
{Dalla Vecchia}, C., {Bower}, R.~G., {Theuns}, T., {Balogh}, M.~L., {Mazzotta},
  P., \& {Frenk}, C.~S. 2004, \mnras, 355, 995


\bibitem[{{De Lucia} {et~al.}(2006){De Lucia}, {Springel}, {White}, {Croton},
  \& {Kauffmann}}]{DeL++06}
{De Lucia}, G., {Springel}, V., {White}, S.~D.~M., {Croton}, D., \&
  {Kauffmann}, G. 2006, \mnras, 366, 499

\bibitem[{{Dickey} \& {Lockman}(1990)}]{D+L90}
{Dickey}, J.~M. \& {Lockman}, F.~J. 1990, \araa, 28, 215

\bibitem[{{Di Matteo} {et~al.}(2003){Di Matteo}, {Allen}, {Fabian}, {Wilson},
  \& {Young}}]{DiM++03}
{Di Matteo}, T., {Allen}, S.~W., {Fabian}, A.~C., {Wilson}, A.~S., \& {Young},
  A.~J. 2003, \apj, 582, 133


\bibitem[{{Di Matteo} {et~al.}(2000){Di Matteo}, {Quataert}, {Allen},
  {Narayan}, \& {Fabian}}]{DiM++00}
{Di Matteo}, T., {Quataert}, E., {Allen}, S.~W., {Narayan}, R., \& {Fabian},
  A.~C. 2000, \mnras, 311, 507


\bibitem[{{Dudik} {et~al.}(2005){Dudik}, {Satyapal}, {Gliozzi}, \&
  {Sambruna}}]{Dud++05}
{Dudik}, R.~P., {Satyapal}, S., {Gliozzi}, M., \& {Sambruna}, R.~M. 2005, \apj,
  620, 113

\bibitem[{{Erwin}(2004)}]{Erw04}
{Erwin}, P. 2004, \aap, 415, 941

\bibitem[{{Fabbiano} \& {Juda}(1997)}]{F+J97}
{Fabbiano}, G. \& {Juda}, J.~Z. 1997, \apj, 476, 666

\bibitem[{{Ferrarese} {et~al.}(2006){Ferrarese}, {C{\^o}t{\'e}}, {Dalla
  Bont{\`a}}, {Peng}, {Merritt}, {Jord{\'a}n}, {Blakeslee}, {Ha{\c s}egan},
  {Mei}, {Piatek}, {Tonry}, \& {West}}]{Fer++06a}
{Ferrarese}, L., {C{\^o}t{\'e}}, P., {Dalla Bont{\`a}}, E., {Peng}, E.~W.,
  {Merritt}, D., {Jord{\'a}n}, A., {Blakeslee}, J.~P., {Ha{\c s}egan}, M.,
  {Mei}, S., {Piatek}, S., {Tonry}, J.~L., \& {West}, M.~J. 2006, \apjl, 644,
  L21

\bibitem[{{Ferrarese} {et~al.}(2006){Ferrarese}, {C{\^o}t{\'e}}, {Jord{\'a}n},
  {Peng}, {Blakeslee}, {Piatek}, {Mei}, {Merritt}, {Milosavljevi{\'c}},
  {Tonry}, \& {West}}]{Fer++06b}
{Ferrarese}, L., {C{\^o}t{\'e}}, P., {Jord{\'a}n}, A., {Peng}, E.~W.,
  {Blakeslee}, J.~P., {Piatek}, S., {Mei}, S., {Merritt}, D.,
  {Milosavljevi{\'c}}, M., {Tonry}, J.~L., \& {West}, M.~J. 2006, \apjs, 164,
  334

\bibitem[{{Ferrarese} \& {Ford}(2005)}]{F+F05}
{Ferrarese}, L. \& {Ford}, H. 2005, Space Science Reviews, 116, 523 (FF05)

\bibitem[{{Ferrarese} \& {Merritt}(2000)}]{F+M00}
{Ferrarese}, L. \& {Merritt}, D. 2000, \apjl, 539, L9

\bibitem[{{Filippenko} \& {Ho}(2003)}]{F+H03}
{Filippenko}, A.~V. \& {Ho}, L.~C. 2003, \apjl, 588, L13

\bibitem[{{Finoguenov} \& {Jones}(2001)}]{F+J01}
{Finoguenov}, A. \& {Jones}, C. 2001, \apjl, 547, L107

\bibitem[{{Fukugita} {et~al.}(1996){Fukugita}, {Ichikawa}, {Gunn}, {Doi},
  {Shimasaku}, \& {Schneider}}]{Fuk++96}
{Fukugita}, M., {Ichikawa}, T., {Gunn}, J.~E., {Doi}, M., {Shimasaku}, K., \&
  {Schneider}, D.~P. 1996, \aj, 111, 1748

\bibitem[{{Garmire} {et~al.}(2000){Garmire}, {Feigelson}, {Broos},
  {Hillenbrand}, {Pravdo}, {Townsley}, \& {Tsuboi}}]{Gar++00}
{Garmire}, G., {Feigelson}, E.~D., {Broos}, P., {Hillenbrand}, L.~A., {Pravdo},
  S.~H., {Townsley}, L., \& {Tsuboi}, Y. 2000, \aj, 120, 1426

\bibitem[{{Gavazzi} {et~al.}(1999){Gavazzi}, {Franzetti}, {Scodeggio},
  {Boselli}, {Pierini}, {Baffa}, {Lisi}, \& {Hunt}}]{Gav++99}
{Gavazzi}, G., {Franzetti}, P., {Scodeggio}, M., {Boselli}, A., {Pierini}, D.,  {Baffa}, C., {Lisi}, F., \& {Hunt}, L.~K. 1999, VizieR Online Data Catalog,
  414, 20065 (Ga99)

\bibitem[{{Gebhardt} {et~al.}(2002){Gebhardt}, {Rich}, \& {Ho}}]{GRH02}
{Gebhardt}, K., {Rich}, R.~M., \& {Ho}, L.~C. 2002, \apjl, 578, L41


\bibitem[{{Gebhardt} {et~al.}(2001){Gebhardt}, {Lauer}, {Kormendy}, {Pinkney},
  {Bower}, {Green}, {Gull}, {Hutchings}, {Kaiser}, {Nelson}, {Richstone}, \&
  {Weistrop}}]{Geb++01}
{Gebhardt}, K., {Lauer}, T.~R., {Kormendy}, J., {Pinkney}, J., {Bower}, G.~A.,
  {Green}, R., {Gull}, T., {Hutchings}, J.~B., {Kaiser}, M.~E., {Nelson},
  C.~H., {Richstone}, D., \& {Weistrop}, D. 2001, \aj, 122, 2469

\bibitem[{{Gebhardt} {et~al.}(2000){Gebhardt}, {Kormendy}, {Ho}, {Bender},
  {Bower}, {Dressler}, {Faber}, {Filippenko}, {Green}, {Grillmair}, {Lauer},
  {Magorrian}, {Pinkney}, {Richstone}, \& {Tremaine}}]{Geb++00}
{Gebhardt}, K., {Kormendy}, J., {Ho}, L.~C., {Bender}, R., {Bower}, G.,
  {Dressler}, A., {Faber}, S.~M., {Filippenko}, A.~V., {Green}, R.,
  {Grillmair}, C., {Lauer}, T.~R., {Magorrian}, J., {Pinkney}, J., {Richstone},  D., \& {Tremaine}, S. 2000, \apjl, 543, L5

\bibitem[{{Geha} {et~al.}(2002){Geha}, {Guhathakurta}, \& {van der
  Marel}}]{GGv02}
{Geha}, M., {Guhathakurta}, P., \& {van der Marel}, R.~P. 2002, \aj, 124, 3073

\bibitem[{{Gehrels}(1986)}]{Geh86}
{Gehrels}, N. 1986, \apj, 303, 336

\bibitem[{{Gerssen} {et~al.}(2002){Gerssen}, {van der Marel}, {Gebhardt},
  {Guhathakurta}, {Peterson}, \& {Pryor}}]{Ger++02}
{Gerssen}, J., {van der Marel}, R.~P., {Gebhardt}, K., {Guhathakurta}, P.,
  {Peterson}, R.~C., \& {Pryor}, C. 2002, \aj, 124, 3270

\bibitem[{{Gilfanov}(2004)}]{Gil04}
{Gilfanov}, M. 2004, \mnras, 349, 146
%


\bibitem[{{Graham} {et~al.}(2001){Graham}, {Erwin}, {Caon}, \&
  {Trujillo}}]{Gra++01}
{Graham}, A.~W., {Erwin}, P., {Caon}, N., \& {Trujillo}, I. 2001, \apjl, 563,
  L11

\bibitem[{{Greene} \& {Ho}(2007)}]{G+H07b}
{Greene}, J.~E. \& {Ho}, L.~C. 2007a, \apj, in press (arXiv0707.2617G)

\bibitem[{{Greene} \& {Ho}(2007)}]{G+H07a}
{Greene}, J.~E. \& {Ho}, L.~C. 2007b, \apj, 667, 131

\bibitem[]{}
H\"aring, N. \& Rix, H.-W., 2004, \apj, 604, L89 


\bibitem[{{Heckman} {et~al.}(2004){Heckman}, {Kauffmann}, {Brinchmann},
  {Charlot}, {Tremonti}, \& {White}}]{Hec++04}
{Heckman}, T.~M., {Kauffmann}, G., {Brinchmann}, J., {Charlot}, S., {Tremonti},
  C., \& {White}, S.~D.~M. 2004, \apj, 613, 109

\bibitem[{{Ho} {et~al.}(2001){Ho}, {Feigelson}, {Townsley}, {Sambruna},
  {Garmire}, {Brandt}, {Filippenko}, {Griffiths}, {Ptak}, \&
  {Sargent}}]{Ho++01}
{Ho}, L.~C., {Feigelson}, E.~D., {Townsley}, L.~K., {Sambruna}, R.~M.,
  {Garmire}, G.~P., {Brandt}, W.~N., {Filippenko}, A.~V., {Griffiths}, R.~E.,
  {Ptak}, A.~F., \& {Sargent}, W.~L.~W. 2001, \apjl, 549, L51

\bibitem[{{Hong} {et~al.}(2005){Hong}, {van den Berg}, {Schlegel}, {Grindlay},
  {Koenig}, {Laycock}, \& {Zhao}}]{Hon++05}
{Hong}, J., {van den Berg}, M., {Schlegel}, E.~M., {Grindlay}, J.~E., {Koenig},
  X., {Laycock}, S., \& {Zhao}, P. 2005, \apj, 635, 907

\bibitem[{{Ichimaru}(1977)}]{Ich77}
{Ichimaru}, S. 1977, \apj, 214, 840

\bibitem[{{Juneau} {et~al.}(2005){Juneau}, {Glazebrook}, {Crampton},
  {McCarthy}, {Savaglio}, {Abraham}, {Carlberg}, {Chen}, {Le Borgne}, {Marzke},
  {Roth}, {J{\o}rgensen}, {Hook}, \& {Murowinski}}]{Jun++05}
{Juneau}, S., {Glazebrook}, K., {Crampton}, D., {McCarthy}, P.~J., {Savaglio},
  S., {Abraham}, R., {Carlberg}, R.~G., {Chen}, H.-W., {Le Borgne}, D.,
  {Marzke}, R.~O., {Roth}, K., {J{\o}rgensen}, I., {Hook}, I., \& {Murowinski},
  R. 2005, \apjl, 619, L135


\bibitem[{{Kollmeier} {et~al.}(2006){Kollmeier}, {Onken}, {Kochanek}, {Gould},
  {Weinberg}, {Dietrich}, {Cool}, {Dey}, {Eisenstein}, {Jannuzi}, {Le Floc'h},
  \& {Stern}}]{Kol++06}
{Kollmeier}, J.~A., {Onken}, C.~A., {Kochanek}, C.~S., {Gould}, A., {Weinberg},
  D.~H., {Dietrich}, M., {Cool}, R., {Dey}, A., {Eisenstein}, D.~J., {Jannuzi},
  B.~T., {Le Floc'h}, E., \& {Stern}, D. 2006, \apj, 648, 128

\bibitem[{{Kormendy} {et~al.}(1997){Kormendy}, {Bender}, {Magorrian},
  {Tremaine}, {Gebhardt}, {Richstone}, {Dressler}, {Faber}, {Grillmair}, \&
  {Lauer}}]{Kor++97}
{Kormendy}, J., {Bender}, R., {Magorrian}, J., {Tremaine}, S., {Gebhardt}, K.,
  {Richstone}, D., {Dressler}, A., {Faber}, S.~M., {Grillmair}, C., \& {Lauer},
  T.~R. 1997, \apjl, 482, L139

\bibitem[{{Kormendy} \& {Richstone}(1995)}]{K+R95}
{Kormendy}, J. \& {Richstone}, D. 1995, \araa, 33, 581

\bibitem[{{Lauer} {et~al.}(2007a){Lauer}, {Faber}, {Richstone}, {Gebhardt},
  {Tremaine}, {Postman}, {Dressler}, {Aller}, {Filippenko}, {Green}, {Ho},
  {Kormendy}, {Magorrian}, \& {Pinkney}}]{Lau++07}
{Lauer}, T.~R., {Faber}, S.~M., {Richstone}, D., {Gebhardt}, K., {Tremaine},
  S., {Postman}, M., {Dressler}, A., {Aller}, M.~C., {Filippenko}, A.~V.,
  {Green}, R., {Ho}, L.~C., {Kormendy}, J., {Magorrian}, J., \& {Pinkney}, J.
  2007a, \apj, 662, 808

\bibitem[{{Lauer} {et~al.}(2007b){Lauer}, {Gebhardt}, {Faber}, {Richstone},
  {Tremaine}, {Kormendy}, {Aller}, {Bender}, {Dressler}, {Filippenko}, {Green},
  \& {Ho}}]{Lau++07}
{Lauer}, T.~R., {Gebhardt}, K., {Faber}, S.~M., {Richstone}, D., {Tremaine},
  S., {Kormendy}, J., {Aller}, M.~C., {Bender}, R., {Dressler}, A.,
  {Filippenko}, A.~V., {Green}, R., \& {Ho}, L.~C. 2007b, \apj, 664, 226

\bibitem[{{Lauer} {et~al.}(1996){Lauer}, {Tremaine}, {Ajhar}, {Bender},
  {Dressler}, {Faber}, {Gebhardt}, {Grillmair}, {Kormendy}, \&
  {Richstone}}]{Lau++96}
{Lauer}, T.~R., {Tremaine}, S., {Ajhar}, E.~A., {Bender}, R., {Dressler}, A.,
  {Faber}, S.~M., {Gebhardt}, K., {Grillmair}, C.~J., {Kormendy}, J., \&
  {Richstone}, D. 1996, \apjl, 471, L79


\bibitem[{{Loewenstein} {et~al.}(2001){Loewenstein}, {Mushotzky}, {Angelini},
  {Arnaud}, \& {Quataert}}]{Loe++01}
{Loewenstein}, M., {Mushotzky}, R.~F., {Angelini}, L., {Arnaud}, K.~A., \&
  {Quataert}, E. 2001, \apjl, 555, L21

\bibitem[{{Lodato} \& {Natarajan}(2006)}]{L+N06}
{Lodato}, G. \& {Natarajan}, P. 2006, \mnras, 371, 1813

\bibitem[{{Madau} \& {Rees}(2001)}]{M+R01}
{Madau}, P. \& {Rees}, M.~J. 2001, \apjl, 551, L27

\bibitem[{{Maoz} {et~al.}(1998){Maoz}, {Koratkar}, {Shields}, {Ho},
  {Filippenko}, \& {Sternberg}}]{Mao++98}
{Maoz}, D., {Koratkar}, A., {Shields}, J.~C., {Ho}, L.~C., {Filippenko}, A.~V.,
  \& {Sternberg}, A. 1998, \aj, 116, 55

\bibitem[{{McNamara} \& {Nulsen}(2007)}]{M+N07}
{McNamara}, B.~R. \& {Nulsen}, P.~E.~J. 2007, \araa, 45, 117

\bibitem[{{Marconi} \& {Hunt}(2003)}]{M+H03}
{Marconi}, A. \& {Hunt}, L.~K. 2003, \apjl, 589, L21

\bibitem[{{Marconi} {et~al.}(2004){Marconi}, {Risaliti}, {Gilli}, {Hunt},
  {Maiolino}, \& {Salvati}}]{Mar++04}
{Marconi}, A., {Risaliti}, G., {Gilli}, R., {Hunt}, L.~K., {Maiolino}, R., \&
  {Salvati}, M. 2004, 222, 49

\bibitem[{{Marconi} {et~al.}}(2007)]{McA++07}
MacArthur, L.~A., Ellis, R.~S., Treu, T., U, V., Bundy, K., Moran S.~M. 2007,
\apj, submitted


\bibitem[{{McLure} \& {Dunlop}(2002)}]{M+D02}
{McLure}, R.~J. \& {Dunlop}, J.~S. 2002, \mnras, 331, 795

\bibitem[{{Mei} {et~al.}(2007){Mei}, {Blakeslee}, {C{\^o}t{\'e}}, {Tonry},
  {West}, {Ferrarese}, {Jord{\'a}n}, {Peng}, {Anthony}, \& {Merritt}}]{Mei++07}
{Mei}, S., {Blakeslee}, J.~P., {C{\^o}t{\'e}}, P., {Tonry}, J.~L., {West},
  M.~J., {Ferrarese}, L., {Jord{\'a}n}, A., {Peng}, E.~W., {Anthony}, A., \&
  {Merritt}, D. 2007, \apj, 655, 144

\bibitem[{{Merloni} \& {Heinz}(2007)}]{M+H07}
{Merloni}, A. \& {Heinz}, S. 2007, \mnras, 381, 589

\bibitem[{{Miller}(2005)}]{Mil05}
Miller, J. 2005, AS\&S, 300, 227-238 



\bibitem[{{Narayan} \& {Yi}(1994)}]{N+Y94}
{Narayan}, R. \& {Yi}, I. 1994, \apjl, 428, L13

\bibitem[{{Pellegrini}(2005)}]{Pel05}
{Pellegrini}, S. 2005, \apj, 624, 155

\bibitem[{{Peng} {et~al.}(2002){Peng}, {Ho}, {Impey}, \& {Rix}}]{Pen++02}
{Peng}, C.~Y., {Ho}, L.~C., {Impey}, C.~D., \& {Rix}, H.-W. 2002, \aj, 124, 266

\bibitem[]{Pet05}
{Peterson}, B.~M. \etal, 2005, \apj, 632, 799

\bibitem[{{Randall} {et~al.}(2004){Randall}, {Sarazin}, \& {Irwin}}]{RSI04}
{Randall}, S.~W., {Sarazin}, C.~L., \& {Irwin}, J.~A. 2004, \apj, 600, 729

\bibitem[{{Rosati} {et~al.}(2002{\natexlab{b}}){Rosati}, {Tozzi}, {Giacconi},
  {Gilli}, {Hasinger}, {Kewley}, {Mainieri}, {Nonino}, {Norman}, {Szokoly},
  {Wang}, {Zirm}, {Bergeron}, {Borgani}, {Gilmozzi}, {Grogin}, {Koekemoer},
  {Schreier}, \& {Zheng}}]{Ros++02}
{Rosati}, P., {Tozzi}, P., {Giacconi}, R., {Gilli}, R., {Hasinger}, G.,
  {Kewley}, L., {Mainieri}, V., {Nonino}, M., {Norman}, C., {Szokoly}, G.,
  {Wang}, J.~X., {Zirm}, A., {Bergeron}, J., {Borgani}, S., {Gilmozzi}, R.,
  {Grogin}, N., {Koekemoer}, A., {Schreier}, E., \& {Zheng}, W.
  2002{\natexlab{b}}, \apj, 566, 667

\bibitem[{{Santra} {et~al.}(2007){Santra}, {Sanders}, \& {Fabian}}]{SSF07}
{Santra}, S., {Sanders}, J.~S., \& {Fabian}, A.~C. 2007, \mnras, 382, 895

\bibitem[{{Sijacki} {et~al.}(2007){Sijacki}, {Springel}, {di Matteo}, \&
  {Hernquist}}]{Sij++07}
{Sijacki}, D., {Springel}, V., {di Matteo}, T., \& {Hernquist}, L. 2007,
  \mnras, 380, 877


\bibitem[{{Sivakoff} {et~al.}(2007){Sivakoff}, {Jord{\'a}n}, {Sarazin},
  {Blakeslee}, {C{\^o}t{\'e}}, {Ferrarese}, {Juett}, {Mei}, \&
  {Peng}}]{Siv++07}
{Sivakoff}, G.~R., {Jord{\'a}n}, A., {Sarazin}, C.~L., {Blakeslee}, J.~P.,
  {C{\^o}t{\'e}}, P., {Ferrarese}, L., {Juett}, A.~M., {Mei}, S., \& {Peng},
  E.~W. 2007, \apj, 660, 1246

\bibitem[{{Sivakoff} {et~al.}(2003){Sivakoff}, {Sarazin}, \& {Irwin}}]{SSI03}
{Sivakoff}, G.~R., {Sarazin}, C.~L., \& {Irwin}, J.~A. 2003, \apj, 599, 218

\bibitem[{{Soldatenkov} {et~al.}(2003){Soldatenkov}, {Vikhlinin}, \&
  {Pavlinsky}}]{SVP03}
{Soldatenkov}, D.~A., {Vikhlinin}, A.~A., \& {Pavlinsky}, M.~N. 2003, Astronomy
  Letters, 29, 298

\bibitem[{{Soltan}(1982)}]{Sol82}
{Soltan}, A. 1982, \mnras, 200, 115


\bibitem[{{Soria} {et~al.}(2006a){Soria}, {Graham}, {Fabbiano}, {Baldi},
  {Elvis}, {Jerjen}, {Pellegrini}, \& {Siemiginowska}}]{Sor++06}
{Soria}, R., {Graham}, A.~W., {Fabbiano}, G., {Baldi}, A., {Elvis}, M.,
  {Jerjen}, H., {Pellegrini}, S., \& {Siemiginowska}, A. 2006a, \apj, 640, 143

\bibitem[{{Soria} {et~al.}(2006b){Soria}, {Graham}, {Fabbiano}, {Baldi},
  {Elvis}, {Jerjen}, {Pellegrini}, \& {Siemiginowska}}]{Sor++06}
{Soria}, R., {Graham}, A.~W., {Fabbiano}, G., {Baldi}, A., {Elvis}, M.,
  {Jerjen}, H., {Pellegrini}, S., \& {Siemiginowska}, A. 2006b, \apj, 640, 126



\bibitem[{{Springel} {et~al.}(2004){Springel}, {White}, \& {Hernquist}}]{SWH04}
{Springel}, V., {White}, S.~D.~M., \& {Hernquist}, L. 2004, 220, 421

\bibitem[{{Sulkanen} \& {Bregman}(2001)}]{S+B01}
{Sulkanen}, M.~E. \& {Bregman}, J.~N. 2001, \apjl, 548, L131

\bibitem[{{Tremaine} {et~al.}(2002){Tremaine}, {Gebhardt}, {Bender}, {Bower},
  {Dressler}, {Faber}, {Filippenko}, {Green}, {Grillmair}, {Ho}, {Kormendy},
  {Lauer}, {Magorrian}, {Pinkney}, \& {Richstone}}]{Tre++02}
{Tremaine}, S., {Gebhardt}, K., {Bender}, R., {Bower}, G., {Dressler}, A.,
  {Faber}, S.~M., {Filippenko}, A.~V., {Green}, R., {Grillmair}, C., {Ho},
  L.~C., {Kormendy}, J., {Lauer}, T.~R., {Magorrian}, J., {Pinkney}, J., \&
  {Richstone}, D. 2002, \apj, 574, 740

\bibitem[{{Treu} {et~al.}(2005){Treu}, {Ellis}, {Liao}, {van Dokkum}, {Tozzi},
  {Coil}, {Newman}, {Cooper}, \& {Davis}}]{Tre++05a}
{Treu}, T., {Ellis}, R.~S., {Liao}, T.~X., {van Dokkum}, P.~G., {Tozzi}, P.,
  {Coil}, A., {Newman}, J., {Cooper}, M.~C., \& {Davis}, M. 2005a, \apj, 633,
  174

\bibitem[{{Treu} {et~al.}(2005){Treu}, {Ellis}, {Liao}, {van Dokkum}}]{Tre++05b}
{Treu}, T., {Ellis}, R.~S., {Liao}, T.~X., {van Dokkum}, 2005b, \apjl, 622, 5



\bibitem[{{Valluri} {et~al.}(2005){Valluri}, {Ferrarese}, {Merritt}, \&
  {Joseph}}]{Val++05}
{Valluri}, M., {Ferrarese}, L., {Merritt}, D., \& {Joseph}, C.~L. 2005, \apj,
  628, 137

\bibitem[{{Volonteri} {et~al.}(2007{\natexlab{a}}){Volonteri}, {Haardt}, \&
  {Gultekin}}]{VHG07}
{Volonteri}, M., {Haardt}, F., \& {Gultekin}, K. 2007{\natexlab{a}}, \mnras\
submitted (arXiv0710.5770)

\bibitem[{{Volonteri} {et~al.}(2007{\natexlab{b}}){Volonteri}, {Lodato}, \&
  {Natarajan}}]{VLN07}
{Volonteri}, M., {Lodato}, G., \& {Natarajan}, P. 2007{\natexlab{b}}, \mnras\
submitted (arXiv0709.0529)

\bibitem[{{Xu} {et~al.}(2005){Xu}, {Xu}, {Zhang}, {Kundu}, {Wang}, \&
  {Wu}}]{Xu++05}
{Xu}, Y., {Xu}, H., {Zhang}, Z., {Kundu}, A., {Wang}, Y., \& {Wu}, X.-P. 2005,
  \apj, 631, 809

\bibitem[{{Woo} \& {Urry}(2002)}]{W+U02}
{Woo}, J.-H. \& {Urry}, C.~M. 2002, \apj, 579, 530


\bibitem[{{Wilson} \& {Yang}(2002)}]{W+Y02}
{Wilson}, A.~S. \& {Yang}, Y. 2002, \apj, 568, 133

\bibitem[{{Zhao} {et~al.}(2005){Zhao}, {Grindlay}, {Hong}, {Laycock}, {Koenig},
  {Schlegel}, \& {van den Berg}}]{Zha++05}
{Zhao}, P., {Grindlay}, J.~E., {Hong}, J.~S., {Laycock}, S., {Koenig}, X.~P.,
  {Schlegel}, E.~M., \& {van den Berg}, M. 2005, \apjs, 161, 429


\end{thebibliography}
\end{document}